\documentclass[%
 reprint,
 amsmath,amssymb,
 aps,
prb,
showkeys,
]{revtex4-2}

\usepackage{graphicx}% Include figure files
\usepackage{dcolumn}% Align table columns on decimal point
\usepackage{bm}% bold math
\usepackage{xcolor} %for colored text
\usepackage{booktabs}
\usepackage[caption=true]{subfig}
\usepackage{float}

\newcommand{ \mb }{ \mathbf } 
\newcommand{ \qq }{ \mb{q}} 
\newcommand{ \kk }{ \mb{k} } 
\newcommand{ \lam }{ \lambda }

\newcommand{ \fdsub }[1]{ f^{0}_{#1} }
\newcommand{ \besub }[1]{ n^{0}_{#1} }
\newcommand{ \velsub }[1]{ \mb{v}_{#1} }

\newcommand{ \gsq }{ \left | g^{smn}_{\kk\qq} \right |^{2}  }
\newcommand{ \elen }[1]{ \epsilon_{#1}  }
\newcommand{ \phen }[1]{ \hbar\omega_{#1}  }
\newcommand{ \Isub }[1]{ \mb{I}_{#1} }
\newcommand{ \Fsub }[1]{ \mb{F}_{#1} }
\newcommand{ \Jsub }[1]{ \mb{J}_{#1} }
\newcommand{ \Gsub }[1]{ \mb{G}_{#1} }
\newcommand{\stirling}[2]{\genfrac{\{}{\}}{0pt}{0}{#1}{#2}}

\newcommand{\ket}[1]{| #1 \rangle}
\newcommand{\bra}[1]{\langle #1 |}

\begin{document}

\preprint{APS/123-QED}

\title{Electron-phonon drag enhancement of transport properties from fully coupled \textit{ab initio} Boltzmann formalism}% Force line breaks with \\
%\thanks{A footnote to the article title}%

\author{Nakib H. Protik}
 \email{nakib@seas.harvard.edu}
\author{Boris Kozinsky}%
 \email{bkoz@seas.harvard.edu}
\affiliation{%
 John A. Paulson School of Engineering and Applied Sciences, Harvard University, Cambridge, Massachusetts 02138, United States.
}%

\date{\today}% It is always \today, today,
             %  but any date may be explicitly specified

\begin{abstract}
We present a combined treatment of the non-equilibrium dynamics and transport of electrons and phonons by carrying out \textit{ab initio} calculations of the fully coupled electron and phonon Boltzmann transport equations. We find that the presence of mutual drag between the two carriers causes the thermopower to be enhanced and dominated by the transport of phonons, rather than electrons as in the traditional semiconductor picture. Drag also strongly boosts the intrinsic electron mobility, thermal conductivity and the Lorenz number. Impurity scattering is seen to suppress the drag-enhancement of the thermal and electrical conductivities, while having weak effects on the enhancement of the Lorenz number and thermopower. We demonstrate these effects in \textit{n}-doped 3C-SiC at room temperature, and explain their origins. This work establishes the roles of microscopic scattering mechanisms in the emergence of strong drag effects in the transport of the interacting electron-phonon gas.
\end{abstract}

%Use showkeys class option if keyword
                              %display desired
\maketitle

\section{Introduction} \label{intro}
In a typical electron (phonon) Boltzmann transport problem, the phonon (electron) system acts as a momentum bath as the latter is assumed to return to equilibrium infinitely fast. This famous ``Bloch's Assumption" \cite{bloch1930elektrischen} was first challenged by Peierls \cite{peierls1930theorie}. Since then, pioneering work by Gurevich \cite{gurevich1989electron} theorized the effect of non-equilibrium phonons and electrons - the mutual drag - on the transport of an interacting electron-phonon gas. Experimental evidence of the phonon drag effect on the thermopower of germanium and silicon was found in the 1950s \cite{frederikse1955thermoelectric, geballe1954seebeck, geballe1955seebeck}. In 1954 Conyers Herring carried out a calculation combining simple analytical models and a partial coupling of the electron and the phonon Boltzmann transport equations (BTEs) \cite{herring1954theory}. To date Herring's analysis of the problem has remained the basis for understanding the drag physics in the context of thermoelectricity. A self-consistent description of the mutual electron and phonon drag effects, however, requires a closed-loop flow of momentum between the two coupled systems of carriers.

To date various approaches have been taken to calculate the electron-phonon mutual drag effect. Some approaches are based on semi-empirical models of interaction and idealized electron and phonon band structures. Approaches in this class include Herring's original work \cite{herring1954theory} on bulk materials and Cantrell and Butcher's work on 2D electron gases \cite{cantrell1987calculation1, cantrell1987calculation2, tsaousidou2001fundamental}. In another approach \cite{mahan2014seebeck}, Mahan, Broido, and Lindsay combined semi-empirical electron-phonon interaction and \textit{ab initio} fitted phonon-phonon interaction with Rode's iterative BTE \cite{rode1970electron} within a partially coupled framework. Very recently, fully \textit{ab initio} methods combining density functional theory (DFT) and the partially coupled BTE were employed by Zhou \textit{et al} \cite{zhou2015ab}, Fiorentini and Bonini \cite{fiorentini2016thermoelectric}, and Macheda and Bonini \cite{ macheda2018magnetotransport}. Lastly, semi-empirical models were combined with the DFT+BTE framework to obtain a solution to the fully coupled electron-phonon (e-ph) BTEs in Ref. \cite{protik2020coupled}.

Here we present for the first time a purely \textit{ab initio} scheme for obtaining the solution of the fully coupled BTEs of the interacting e-ph gas. We apply this method to the $\textit{n}$-doped cubic phase of silicon carbide (3C-SiC), which is a large band gap material widely used in thermoelectrics, power electronics and quantum computing. We calculate the effect of drag on the various transport coefficients with and without the presence of charged impurity scattering and interpret the results in terms of the various electron-phonon scattering processes. At room temperature and over a wide range of the carrier concentrations we find that there is a surprisingly strong drag-driven increase of (i) the electron mobility in the absence of impurities and (ii) the thermopower and the Lorenz factor with and without impurity scattering. The result is surprising because strong drag behavior is typically  associated with low temperatures. Our results build on the recent formulation of the coupled e-ph BTEs and prediction of strong phonon drag gain of electron mobility in GaAs using semi-empirical models for e-ph scattering \cite{protik2020coupled}. In this work the e-ph matrix elements are calculated completely from first principles. This allows us to capture the full wave-vector dependence of the e-ph coupling, which is absent in simpler analytical models. This method accurately captures the details of the electron and phonon band structures that semi-empirical models cannot. Moreover, since the method is \textit{ab initio}, it enables the study of materials for which the semi-empirical model parameters are not known.

\section{Theory and computation} \label{transport}
\subsection{Coupled electron-phonon transport theory}
Here we present the coupled e-ph BTEs originally formulated in Ref. \cite{protik2020coupled}. Within the weak-field, linear response regime, the electron and phonon distribution functions are, respectively,
\begin{align}\label{eq:distrib}
    f_{m\mathbf{k}} &\approx f^{0}_{m\mathbf{k}}\left[ 1 + (1 - f^{0}_{m\mathbf{k}})\Psi_{m\mathbf{k}} \right] \nonumber \\
    n_{s\mathbf{q}} &\approx n^{0}_{s\mathbf{q}}\left[ 1 + (1 + n^{0}_{s\mathbf{q}})\Phi_{s\mathbf{q}} \right],
\end{align}
where $m$ ($s$) denotes the electron (phonon) band (branch), $\mathbf{k}$ ($\mathbf{q}$) is the electron (phonon) wave vector, $f^{0}_{m\mathbf{k}}$ ($n^{0}_{s\mathbf{q}}$) is the Fermi-Dirac (Bose-Einstein) distribution function, and $\Psi_{m\mathbf{k}}$ ($\Phi_{s\mathbf{q}}$) measures the deviation of the electron (phonon) from equilibrium due to the presence of an external field.

The deviation functions themselves can be written as follows:
\begin{align}\label{eq:dev}
    \Psi_{m\mathbf{k}} &= -\beta\nabla T\cdot \mathbf{I}_{m\mathbf{k}} -\beta\mathbf{E}\cdot \mathbf{J}_{m\mathbf{k}} \nonumber \\
    \Phi_{s\mathbf{q}} &= -\beta\nabla T\cdot \mathbf{F}_{s\mathbf{q}} -\beta\mathbf{E}\cdot \mathbf{G}_{s\mathbf{q}},
\end{align}
where $\beta \equiv (k_{\text{B}}T)^{-1}$, $T$ is the temperature, $k_{\text{B}}$ is the Boltzmann constant, $\nabla T$ is the temperature gradient field, $\mathbf{E}$ is the electric field, $\mathbf{I}_{m\mathbf{k}}$ ($\mathbf{F}_{s\mathbf{q}}$) are the response coefficients of the electron (phonon) states to the temperature gradient, and $\mathbf{J}_{m\mathbf{k}}$ ($\mathbf{G}_{s\mathbf{q}}$) are the response coefficients of the electron (phonon) states to the electric field.

The coupled BTEs for the temperature gradient and the electric field, respectively, can then be written succinctly as \cite{protik2020coupled}
\begin{align}\label{eq:cbtes}
%\nabla T: \nonumber \\
\mathbf{I} &= \mathbf{I}^{0} + \Delta\mathbf{I}^{\text{S}}[\mathbf{I}] + \Delta\mathbf{I}^{\text{D}}[\mathbf{F}] \nonumber \\
\mathbf{F} &= \mathbf{F}^{0} + \Delta\mathbf{F}^{\text{S}}[\mathbf{F}]
+ \Delta\mathbf{F}^{\text{D}}[\mathbf{I}] \nonumber \\ 
%\mathbf{E}: \nonumber \\
\mathbf{J} &= \mathbf{J}^{0} + \Delta\mathbf{J}^{\text{S}}[\mathbf{J}] + \Delta\mathbf{J}^{\text{D}}[\mathbf{G}] \nonumber \\
\mathbf{G} &= \Delta\mathbf{G}^{\text{S}}[\mathbf{G}] + \Delta\mathbf{G}^{\text{D}}[\mathbf{J}],
\end{align}
where we have dropped the indices to avoid clutter.

In Eq. \ref{eq:cbtes}, the terms with the superscript nought describe drift due to the applied field. The (out-)scattering rates that enter this term are equivalent to the inverse lifetimes obtained from the leading order self-energy diagrams in the Migdal approximation. The expressions for $\mathbf{I^{0}}$, $\mathbf{F^{0}}$, and $\mathbf{J^{0}}$ are given by Eqs. \ref{eq:I0}, \ref{eq:F0}, and \ref{eq:J0}, respectively. The scattering rates entering these quantities are given by Eqs. \ref{eq:We} and \ref{eq:Wph}. The electric field phonon BTE does not contain such a field term because phonons do not carry charge. The terms with superscript S (for ``self") are functionals of the deviation function of the same carrier species whereas the terms with superscipt D (for ``drag") are functionals of the deviation function of the other species. These terms, given by Eqs. \ref{eq:IS}, \ref{eq:ID}, \ref{eq:FS}, \ref{eq:FD}, \ref{eq:JS}, \ref{eq:JD}, \ref{eq:GS}, and \ref{eq:GD}, describe the in-scattering corrections and the momentum exchange between the two interacting gases that are not included in the field terms. In the absence of the self and drag terms, one recovers the widely used relaxation time approximation (RTA). Specifically, by phonon (electron) drag we mean the transfer of momentum via scattering from the phonons (electrons) to the electrons (phonons). In this work we obtain the full solution of the coupled set of BTEs by starting with the RTA solution and then iterating to self-consistency. From the final solution of the coupled BTEs we compute the phonon and electronic components of the thermal conductivity, electron mobility, and the thermopower. In this work we calculate both the electron-phonon and the phonon-phonon matrix elements from first principles.

The \textit{ab initio} electron-phonon matrix elements are given by \cite{ponce2016epw}
\begin{equation}\label{eq:g}
    g_{\kk\qq}^{smn} = \sqrt{\dfrac{\hbar}{2\omega_{s\mb{q}}}} \bra{\zeta_{n[\mb{k}+\qq]}}\nabla_{s\qq}V_{\text{SCF}}\ket{\zeta_{m\mb{k}}},
\end{equation}
where $\hbar$ is the reduced Planck's constant, $\ket{\zeta_{m\mb{k}}}$ is the Kohn-Sham state, and $\nabla_{s\qq}V_{\text{SCF}}$ is the derivative of the self-consistent Kohn-Sham potential. Note that the derivative is atomic mass normalized. In the Appendix we provide the matrix elements and the corresponding RTA scattering rates expressions using simple analytical models of e-ph interaction. This provides a valuable check for the \textit{ab initio} calculations.

The \textit{ab initio} phonon-phonon matrix elements are given by \cite{li2014shengbte}
\begin{equation}
V^{\pm}_{\lambda\lambda'\lambda''} = \sum_{<i>jk}\sum_{\alpha\beta\gamma}\Psi^{\alpha\beta\gamma}_{ijk}\dfrac{e^{i\alpha}_{s}(\qq)e^{j\beta}_{s'}(\pm\mathbf{q}')e^{k\gamma}_{s''}(-\mathbf{q}'')}{\sqrt{m_{i}m_{j}m_{k}}},
\end{equation}
where the phonon state $\lambda \equiv (s,\mb{q})$, Greek indices denote the Cartesian directions, $i, j, k$ identify atoms in the supercell, the triangular brackets denote restricted summation over the atoms in the primitive unit cell, $\Psi$ is the third-order force constants (IFC3) tensor, $m_{i}$ denotes the mass of atom $i$, and $e_{s}^{i}(\qq)$ denotes the branch $s$ phonon eigenfunction in the cell where atom $i$ resides for wave vector $\qq$.

Once the solutions to Eqs. \ref{eq:cbtes} are known, one can calculate the full set of transport coefficient tensors. From the electronic charge current we get
\begin{equation}
\begin{Bmatrix}
\sigma^{\alpha\beta} \\
[\sigma S]^{\alpha\beta}
\end{Bmatrix}
=\dfrac{2e}{Vk_{B}T}\sum_{\nu}f^{0}_{\nu}(1-f^{0}_{\nu})v^{\alpha}_{\nu}\times 
\begin{Bmatrix}
J^{\beta}_{\nu} \\
-I^{\beta}_{\nu}
\end{Bmatrix}
%\stirling{\sigma^{\alpha\beta}}{[\sigma S]^{\alpha\beta}} = \dfrac{2e}{Vk_{B}T}\sum_{\nu}f^{0}_{\nu}(1-f^{0}_{\nu})v^{\alpha}_{\nu}\times \stirling{J^{\beta}_{\nu}}{-I^{\beta}_{\nu}},
\end{equation}
where $e$ is the magnitude of the electronic charge, $V$ is the supercell volume, $v^{\alpha}_{\nu}$ is the velocity of the electron in state $\nu \equiv (m,\mb{k})$ in the Cartesian direction $\alpha$, $\sigma$ is the electronic conductivity tensor, and $S = Q$ is the Seebeck thermopower tensor. The mobility tensor is defined as $\mu = \sigma(ne)^{-1}$, where $n$ is the charge carrier concentration.

From the electronic heat current we get
\begin{equation}
\begin{Bmatrix}
\alpha^{\alpha\beta}_\text{el} \\
\kappa^{\alpha\beta}_{0,\text{el}}
\end{Bmatrix}
= -\dfrac{2}{Vk_{B}T}\sum_{\nu}(\epsilon_{\nu}-E_{\text{F}})f^{0}_{\nu}(1-f^{0}_{\nu})v^{\alpha}_{\nu}\times
\begin{Bmatrix}
J^{\beta}_{\nu} \\
-I^{\beta}_{\nu}
\end{Bmatrix}
%\stirling{\alpha^{\alpha\beta}_\text{el}}{\kappa^{\alpha\beta}_{0,\text{el}}} = -\dfrac{2}{Vk_{B}T}\sum_{\nu}(\epsilon_{\nu}-E_{\text{F}})f^{0}_{\nu}(1-f^{0}_{\nu})v^{\alpha}_{\nu}\times\stirling{J^{\beta}_{\nu}}{-I^{\beta}_{\nu}},
\end{equation}
where $E_{\text{F}}$ is the chemical potential,  $\alpha_\text{el}$ is related to the electronic Peltier thermopower tensor $Q_\text{el} = \alpha_\text{el} (\sigma T)^{-1}$ and $\kappa_{0,\text{el}}$ is the electronic thermal conductivity tensor in the zero electric field condition. The electronic thermal conductivity in the zero current condition is given by $\kappa_{\text{el}} = \kappa_{0,\text{el}} - \alpha_\text{el}Q$.

From the phonon heat current equation we get
\begin{equation}
\begin{Bmatrix}
\alpha^{\alpha\beta}_{\text{ph}}\\
\kappa^{\alpha\beta}_{\text{ph}}
\end{Bmatrix}
= \dfrac{1}{Vk_{B}T}\sum_{\lam}\hbar\omega_{\lam}n^{0}_{\lam}(1+n^{0}_{\lam})v^{\alpha}_{\lam}\times
\begin{Bmatrix}
-G^{\beta}_{\lam}\\
F^{\beta}_{\lam}
\end{Bmatrix}
%\stirling{\alpha^{\alpha\beta}_{\text{ph}}}{\kappa^{\alpha\beta}_{\text{ph}}} = \dfrac{1}{Vk_{B}T}\sum_{\lam}\hbar\omega_{\lam}n^{0}_{\lam}(1+n^{0}_{\lam})v^{\alpha}_{\lam}\times\stirling{-G^{\beta}_{\lam}}{F^{\beta}_{\lam}},
\end{equation}
where $v^{\alpha}_{\lambda}$ is the velocity of the phonon in state $\lambda$ in the Cartesian direction $\alpha$, $\alpha_{\text{ph}}$ is related to the phonon Peltier thermopower tensor $Q_{\text{ph}} = \alpha_{\text{ph}}(\sigma T)^{-1}$ and $\kappa_{\text{ph}}$ is the phonon thermal conductivity tensor.

Lastly, the Lorenz number is defined as $L = \kappa_{\text{el}}(\sigma T)^{-1}$.

For cubic systems, such as 3C-SiC, all the off-diagonal components of each transport tensor are zero and all the diagonal components are identical.\\

\subsection{Computational details}
We use the \texttt{Quantum Espresso} \cite{giannozzi2017advanced} suite for our density functional theory and density functional perturbation theory calculations. The norm conserving, Perdew-Zunger (local density approximation) pseudopotential \cite{perdew1981self} is used. Our calculated relaxed lattice constant is $4.339$ A. This is in good agreement with the experimental value of $4.360$ A in Ref. \cite{smiltens1960silicon} and previously calculated values of $4.34$ A, $4.33$ A, and $4.342$ A in Refs. \cite{lindsay2013ab}, \cite{katre2017exceptionally}, and \cite{wang2017strong}, respectively. We interface with the \texttt{EPW} code \cite{ponce2016epw, giustino2007electron, verdi2015frohlich} to compute the electron-phonon matrix elements. We use a $6\times 6\times 6$ coarse electron (\textbf{k}) and phonon (\textbf{q}) wave vector mesh. We print out the information related to the Wannier representation of the electron-phonon matrix element from \texttt{EPW} and read into our transport code. The polar Wannier to Bloch transformation of the matrix elements is performed and the data is saved for reuse for the concentration sweep and the various types of BTE solutions. The computational cost is reduced by using an effective transport window of $0.4$ eV from the conduction band minimum. Doping is simulated by moving the chemical potential and the chosen energy window is sufficiently large for the concentration range considered in this work. All results presented in the text are calculated on a $65\times 65\times 65$ \textbf{q} mesh and $130\times 130\times 130$ \textbf{k}, \textbf{k}+\textbf{q} mesh. This means that in the electronic transition probabilities calculation, the electronic and the phonon wave vector meshes are both the $130\times 130\times 130$. All the phonon transition probabilities are calculated on the $65\times 65\times 65$ mesh, where the summation over the electronic states are performed over the finer $130\times 130\times 130$ mesh. During the iterative solution of the coupled Boltzmann transport equations (BTEs), the phonon quantities are interpolated onto the finer mesh whenever needed. We use the short-hand $(65, 130)$ for this choice of the meshes.
In the Appendix we discuss code validation and numerical convergence.
The scattering transition probabilities are given by Eqs. \ref{eq:X}, \ref{eq:Y}, and \ref{eq:W}. These require evaluation of the energy conserving delta functions for the elecron-phonon and phonon-phonon processes. The analytic tetrahedron method \cite{lambin1984computation} is used to approximate these delta functions. Electron-impurity scattering (assuming singly charged dopants) is calculated using the Brooks-Herring model \cite{chattopadhyay1981electron}. In this model, the impurity is taken to be a static, Yukawa-type scattering center and the electron-defect scattering is treated in the first Born approximation. The phonon-phonon matrix elements are calculated from the real space third order force constants (IFC3s). The IFC3s are calculated using finite displaced supercells. A $5\times 5\times 5$ ($250$ atoms) supercell is used with a $6$ nearest neighbor ($0.548$ nm) cut-off. The \texttt{thirdorder.py} \cite{li2012thermal, li2014shengbte} code is used to generate the displaced supercells. \texttt{Quantum Espresso} is used to compute the forces in the $292$ displaced supercells. The \texttt{thirdorder.py} code is then used to read the forces and compute the IFC3s, which are then used as an input to the transport code. Phonon-isotope scattering is calculated in the Tamura model \cite{tamura1983isotope}.\\

\section{Results and discussion} \label{results}
\subsection{Scattering rates}
\begin{figure}%[H]
    \centering
	\subfloat{\includegraphics[scale=0.423]{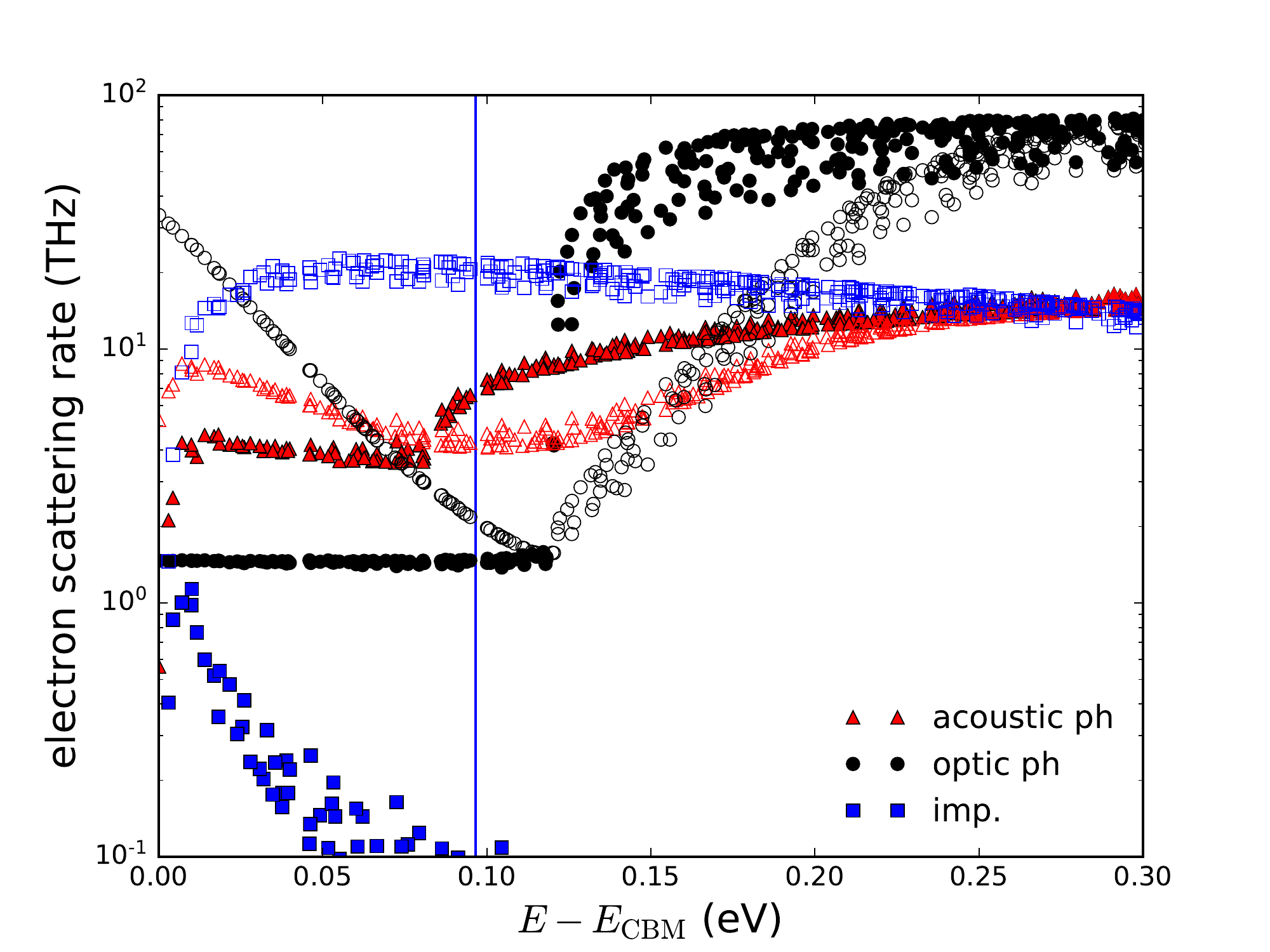}}
	\caption{Breakdown of the electronic RTA scattering rates for carrier concentrations $10^{16}$cm$^{-3}$ (solid symbols) and $10^{20}$cm$^{-3}$ (hollow symbols) at $300$ K. For the latter, the chemical potential is shown by the blue vertical line.}
	\label{fig:eph}
\end{figure}

The electronic scattering rates in the RTA do not include drag, but are already useful for the physical interpretation of the roles of the various scattering mechanisms in transport. Fig. \ref{fig:eph} shows the RTA rates for the $n$-doped 3C-SiC at $300$ K. The zero of the energy axis is at the conduction band minimum energy, $E_\text{CBM}$. In the low doping case, when the electron chemical potential is in the band gap, the low energy electrons predominantly scatter against low energy acoustic phonons via the piezoelectric and acoustic deformation potential type interactions. Both these interactions are quasielastic in nature since the phonon energies are small compared to the electron energies. Around $80$ meV, the acoustic scattering rate increases sharply. This originates from the deformation potential type scattering of the longitudinal acoustic (LA) phonon at the $X$-point of the Brillouin zone. Note here that while the acoustic and optic vibrational modes are distinct near the zone center, the same is not true near the zone boundary. Moreover, the zone boundary LA phonon energy is close to that of the optic phonons. As such, the deformation potential coupling of the electrons to the $X$-point LA phonon is optical in nature. In Ref. \cite{patrick1966high} it has been shown by group theoretic analysis that since 3C-SiC has a three-fold degenerate conduction band minimum at the $X$-point, the phase space for inter-valley scattering is severely restricted and only the LA phonons at the $X$ point can contribute significantly. Thus, the kink at $80$ meV is due to the onset of the inter-valley $X$-point LA phonon emission. Note that this interaction is inelastic since the phonon energy is not negligible. Similarly, the longitudinal optic (LO) phonon emission onset can be clearly seen around $120$ meV. Polar optical phonon scattering dominates at higher energies. Very similar scattering rates features have been reported in Ref. \cite{meng2019phonon}. For the high doping case (chemical potential in the conduction band), there is a large dip in the inelastic polar LO phonon scattering rates. This is a phase-space reduction effect. The scattering phase space is essentially the RTA scattering rates expression Eq. \ref{eq:We} with the e-ph matrix elements set to unity. As such, both the energy conservation effect (conveyed by the delta function) and the statistics effect (conveyed by the presence of the Fermi-Dirac functions) are included in the definition of the phase space. Physically, when approaching the chemical potential from below, the electronic occupations number sharply decreases, leading to a reduction in the LO phonon absorption rates. On the other hand, when approaching the chemical potential from above, the lower energy states are nearly full, causing the LO phonon emission rates to decrease. The combined effect is that the electrons near the chemical potential interact less with LO phonons, which again makes the quasielastic low-energy phonon and inelastic high-energy acoustic phonon scattering the dominant channels.
Thus, in both the low and high doping regimes, the transport active electrons pump momentum into the low energy and zone boundary acoustic and the LO phonons, rendering these phonons strongly drag active. Note also that at high doping concentrations, the charged impurity scattering channel dominates the electron-phonon channel. 

\begin{figure}%[H]
    \centering
	\includegraphics[scale=0.423]{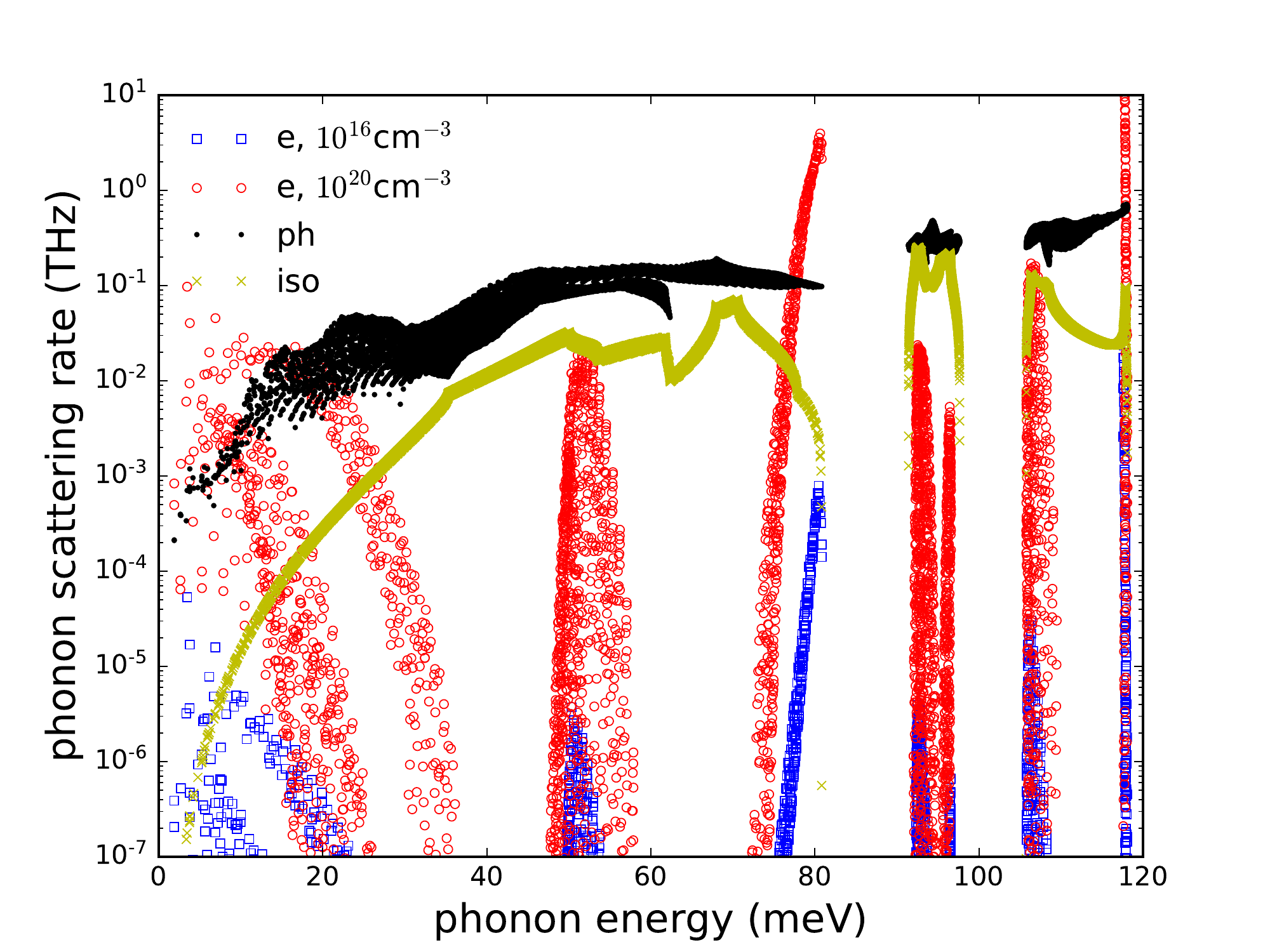}
	\caption{Breakdown of the phonon RTA scattering rates at $300$ K into phonon-(e)lectron, phonon-(ph)onon, and phonon-(iso)tope channels.}
	\label{fig:ph}
\end{figure}

Fig. \ref{fig:ph} shows the phonon RTA scattering rates breakdown into the phonon-electron, phonon-isotope, and phonon-phonon channels. %The phonon-isotope scattering rates are calculated using the Tamura formula \cite{tamura1983isotope}. 
We do not include grain boundary scattering in this work. The phonon-phonon scattering rates increase with phonon energy. The phonon-isotope scattering rates are weak for low energy phonons, but are comparable to the phonon-phonon rates for near zone boundary acoustic phonons and the optic phonons. The phonon-electron scattering rates for low energy acoustic phonons drop of sharply with increasing energy, which is typical of piezoelectric and acoustic deformation type scattering. There is strong scattering at $80$ meV, corresponding to optical deformation type scattering with the $X$-point LA phonon, which gives dominant contribution to intervalley scattering as mentioned earlier. The phonon-electron scattering rates for the $120$ meV LO phonons are also strong owing to the polar nature of their coupling to the electrons. The momentum received from the electrons can be distributed and dissipated into the phonon system via anharmonic phonon-phonon interaction and fed back into the electron system via phonon-electron interaction. In general, the flow of momentum back into the electron system results in an enhancement of the electronic transport coefficients (mobility, thermal conductivity, and thermopower) due to phonon drag. On the other hand, a low overall rate of momentum dissipation within the phonon system manifests itself as electron drag induced enhancement of the phonon transport coefficients (thermal conductivity and thermopower), given that the phonon system has received excess momentum from the electron system. Note that the low energy acoustic phonons have low anharmonic scattering rates and have fewer momentum destroying Umklapp anharmonic scattering. As such, they can sustain the momentum received from the electronic system for a long time in contrast to the shorter lived optical phonons. 

While the analysis presented above based on the RTA scattering rates provides a relatively simple qualitative picture, the iteration process of the coupled BTEs non-trivially mixes the momentum in the interacting system of electrons and phonons. Nonetheless, the RTA scattering rates allow us to interpret the drag phenomena predicted by the self-consistent coupled solutions.

\subsection{Thermal conductivity}
\begin{figure}%[H]
    \centering
	\includegraphics[scale=0.423]{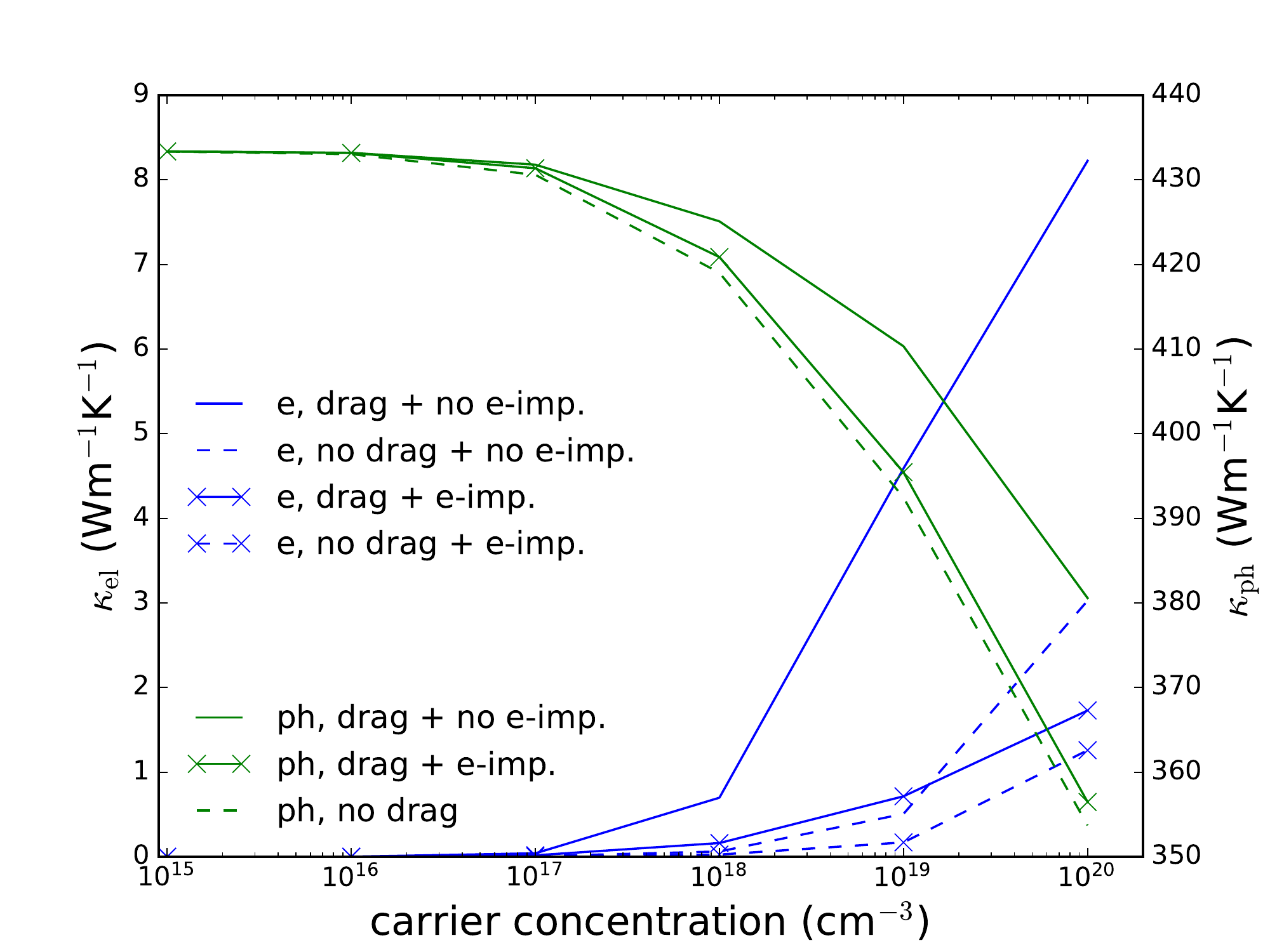}
	\caption{Phonon (right axis, green) and electron (left axis, blue) thermal conductivity as a function of carrier concentration at $300$ K. Note that for the no drag case for phonons (dashed green line), the impurity scattering channel in the electronic system is irrelevant for phonon transport.}
	\label{fig:kappa}
\end{figure}

\begin{figure}%[H]
    \centering
	\includegraphics[scale=0.423]{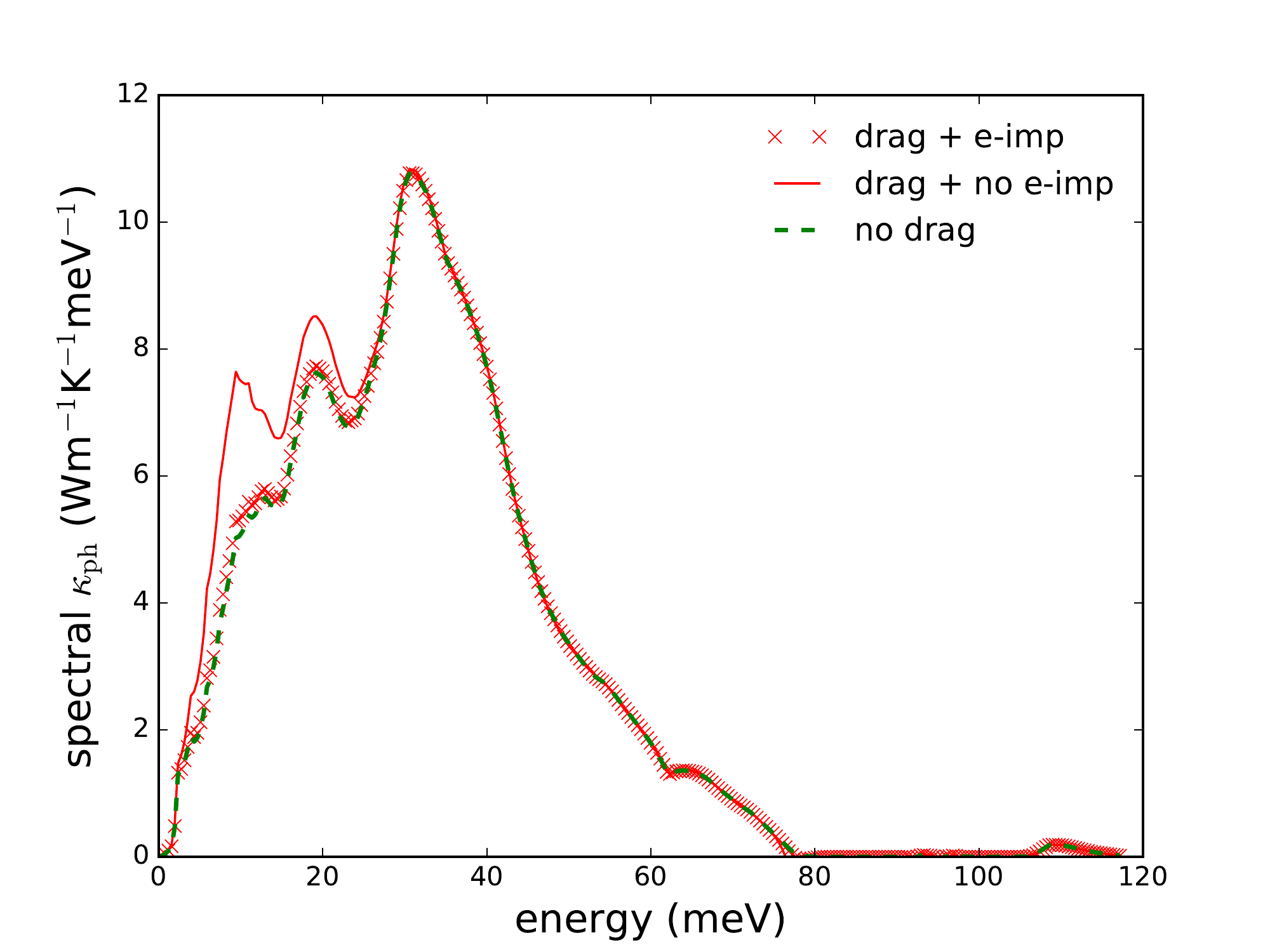}
	\caption{Spectral decomposition of the phonon thermal conductivity for a carrier concentration of $10^{20}$ cm$^{-3}$.}
	\label{fig:speckappa_ph}
\end{figure}

Fig. \ref{fig:kappa} shows the thermal conductivity, $\kappa$, as a function of the carrier concentration. The phonon contribution completely dominates the electronic contribution to $\kappa$ over the entire range of carrier concentrations. In the low doping limit, the computed phonon $\kappa$ ($433$ Wm$^{-1}$K$^{-1}$) is similar to the calculated values in literature - about $10$\% lower than those reported in Refs. \cite{lindsay2013ab}, \cite{katre2017exceptionally} and \cite{wang2017strong}. The literature calculations are formally equivalent to the decoupled, iterative phonon BTE calculation in our formulation. The effect of drag on the phonon $\kappa$, while increasing with carrier concentration, is overall small. This is a consequence of the fact that the drag active zone center and zone boundary acoustic and high energy optical phonons contribute weakly to the phonon $\kappa$. %The spectral phonon $\kappa$ is given in Appendix A section \textit{Spectral transport coefficients} to further demonstrate this point. 
At $10^{20}$ cm$^{-3}$ doping concentration, the electron drag induced gain of the phonon $\kappa$ is only around $8$\%, if charged impurity scattering of electrons is not included. If charged impurity scattering is included, this drag gain is uniformly negligible, since the amount of momentum feedback from the electronic system diminishes with increasing carrier concentration due to the increasing dissipation of electronic momentum by charged impurities. To understand this effect we show in Fig. \ref{fig:speckappa_ph} the spectral phonon $\kappa$ for a carrier concentration of $10^{20}$ cm$^{-3}$. The contribution to thermal conductivity comes from a large range of acoustic phonon frequencies, which is a consequence of the fact that in this material, the LA phonons can sustain a high velocity over a large portion of the Brillouin zone. Zone center acoustic phonons do not contribute strongly to the thermal conductivity since the specific heat for those modes are small. Similarly, optic phonons and zone boundary acoustic phonons contribute negligibly due to their low group velocities. The dashed green curve gives the spectral $\kappa$ when the phonon BTE is decoupled from the electron BTE, i.e. when there is no electron drag effect. Comparing to the solid red curve denoting the case where drag is included but neglecting electron-charged impurity scattering, we see that the effect of drag is to boost the thermal conductivity contribution from acoustic phonons between around $5$ and $25$ meV. These are precisely those phonons that have higher or comparable phonon-electron scattering rates compared to the phonon-phonon channel. Moreover, these small energy phonons are also the ones into which the electrons pump momentum strongly, setting up a robust circulation of momentum between the two systems. Thus, these strongly drag active phonons dissipate momentum less when drag is considered as opposed to when it is neglected. However, since the additional gain in $\kappa$ is small compared to the contribution from the whole acoustic phonon spectrum, the overall gain in $\kappa$ due to drag is modest. The red crosses denote the case where drag is included along with the charged impurity scattering channel for electrons. In this case, the effect of drag is destroyed since the momentum transferred by the acoustic phonons to the electrons is dissipated by strong impurity scattering which is the dominant scattering mechanism at high carrier concentrations. The calculated weak electron drag effect on the phonon $\kappa$ is in agreement with the findings in Ref. \cite{protik2020coupled} for GaAs and validates the fact that numerous phonon $\kappa$ calculations on different materials that have ignored the electron drag effect have, nevertheless, found good agreement with experiments. 

The electronic contribution to $\kappa$, while negligible compared to the phonon counterpart, features strong phonon drag effect at high carrier concentrations when charged impurity scattering is ignored for the same reasons given above -- in the absence of the dissipative impurity scattering channel, there is a persistent circulation of momentum between the electron and phonon systems mediated by the interaction of electrons with strongly drag active zone center and zone boundary acoustic and high energy optical phonons. At $10^{20}$ cm$^{-3}$ doping concentration, the phonon drag gain of the electronic $\kappa$ is $37$\% in the presence of charged impurity electron scattering, and $171$\% in the absence of impurities.

\subsection{Carrier mobility}
\begin{figure}%[H]
    \centering
	\includegraphics[scale=0.423]{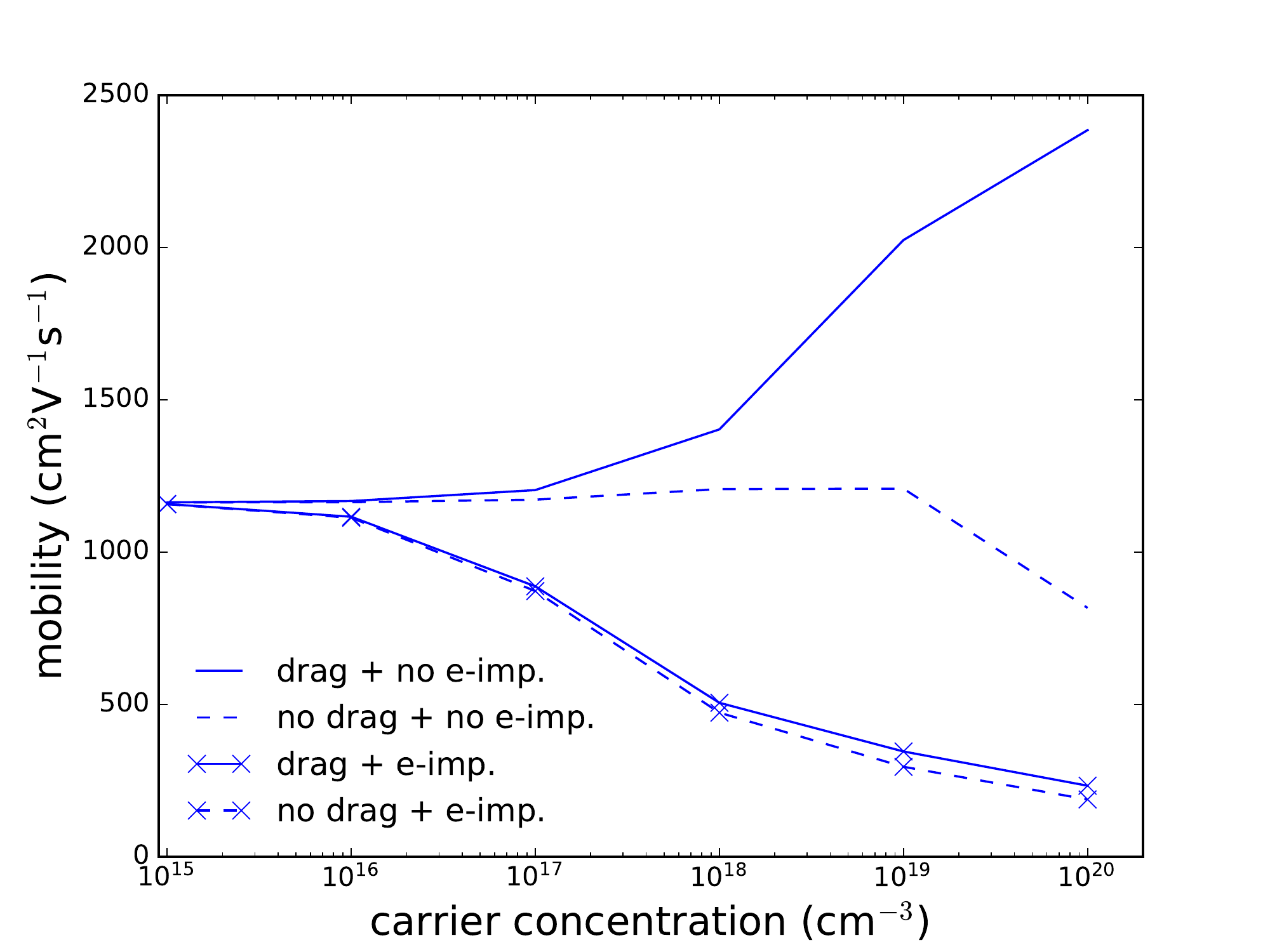}
	\caption{Electron mobility as a function of carrier concentration at $300$ K.}
	\label{fig:mu}
\end{figure}

Fig. \ref{fig:mu} shows the electron mobility versus carrier concentration. The charged impurity scattering channel begins to limit the mobility above $10^{16}$ cm$^{-3}$ carrier concentration. The highest measured mobility at room temperature is $980$ cm$^{2}$V$^{-1}$s$^{-1}$ for a carrier concentration of $4\times10^{16}$ cm$^{-3}$ \cite{nelson1966growth}, which is in excellent agreement with our calculated values of $1116$ and $888$ cm$^{2}$V$^{-1}$s$^{-1}$ at $10^{16}$ and $10^{17}$ cm$^{-3}$, respectively, when charge impurity scattering is included in the calculation. %For the same reasons as for electronic thermal conductivity, the phonon drag enhancement of the mobility increases with increasing carrier concentration. 
In the absence of impurity scattering of electrons, the phonon drag gain of the mobility is substantial - $16$\% ($191$\%) at $10^{18}$ ($10^{20}$) cm$^{-3}$ carrier concentration. If techniques such as modulated doping can be realized on bulk samples, then our prediction of the strong phonon drag gain of mobility can be experimentally tested.

\begin{figure}%[H]
    \centering
	\includegraphics[scale=0.423]{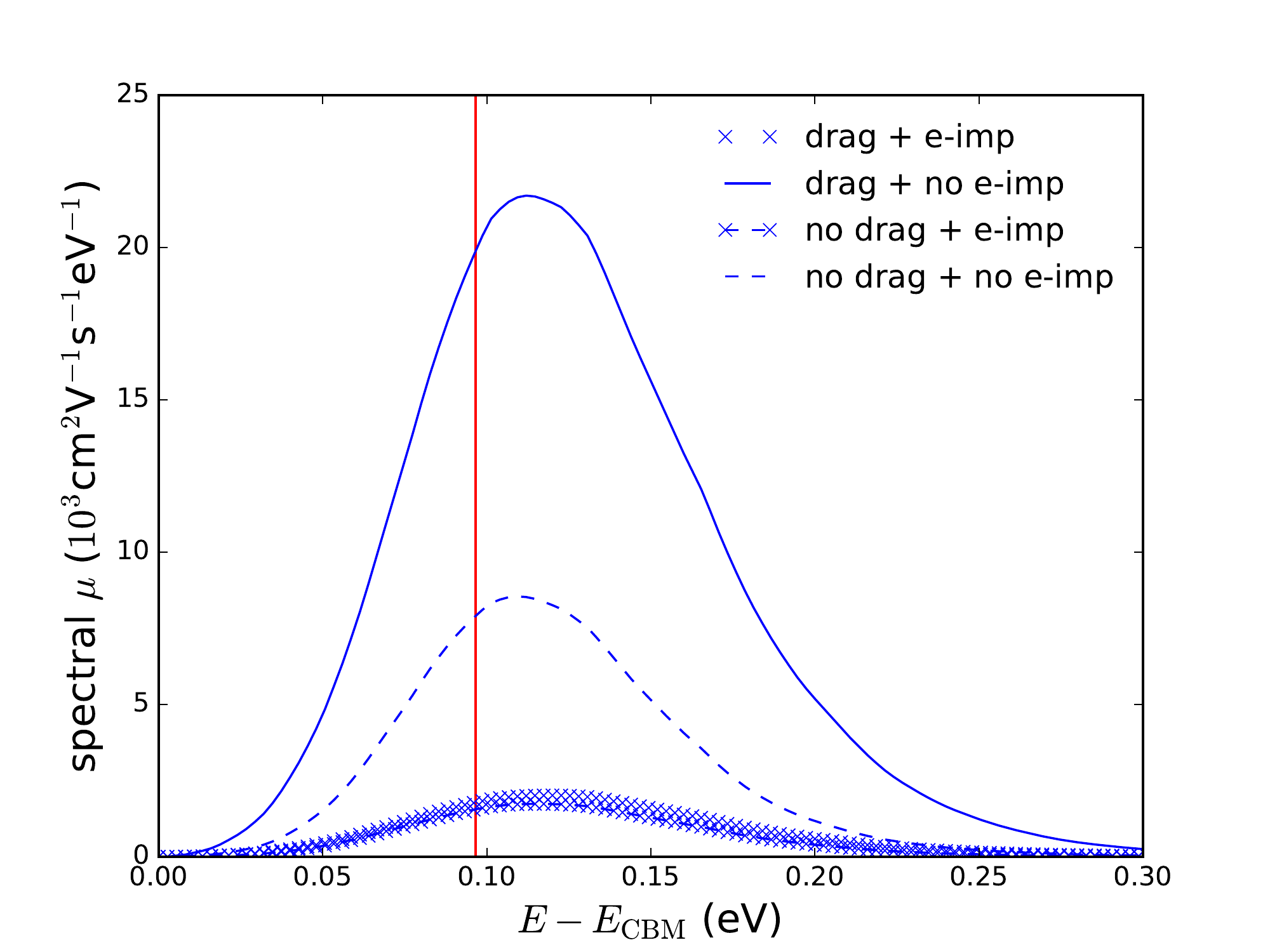}
	\caption{Spectral decomposition of the electron mobility for a carrier concentration of $10^{20}$ cm$^{-3}$. The zero of the energy axis is at the conduction band minimum (CBM). The vertical red line denotes the position of the chemical potential.}
	\label{fig:specmob}
\end{figure}

To demonstrate the drag effect we present in Fig. \ref{fig:specmob} the spectrum of the electronic mobility $\mu$ for a carrier concentration of $10^{20}$ cm$^{-3}$. The vertical red line denotes the position of the chemical potential at this concentration. Firstly, the peak of the mobility contribution is roughly centered around the chemical potential. This is expected for two reasons: (1) the phase-space reduction of inelastic scattering of electrons from optic and zone boundary acoustic phonons happens near the chemical potential, and (2) the Fermi window function $f^{0}(1 - f^{0})$ that appears in the mobility expression peaks at the chemical potential. The small asymmetry in the spectrum is due to the energy dependence of the electronic density of states which roughly goes as $\sqrt{E - E_{\text{CBM}}}$ and the fact that the scattering rates themselves have energy dependence. The solid (dashed) blue curve denotes the case with (without) phonon drag while ignoring electron-charged impurity scattering. There is a large drag boost in the spectral mobility coming from a large energy range. This is a consequence of the fact that in this material, electrons interact strongly with the zone center and zone boundary acoustic phonons and the polar optic phonons. In turn, for these phonon modes, the phonon-electron scattering rates dominate the phonon-phonon rates. As such, a robust momentum circulation is sustained between the electron and the phonon systems mediated via these phonon modes. Thus, electrons dissipate significantly less momentum when drag is included as opposed to when it is ignored. On the other hand, when the strongly dissipative electron-charged impurity scattering channel is turned on, the drag effect is destroyed. This is seen by comparing the blue crosses (drag with electron-impurity scattering) and blue cross-dashed (no drag with electron-impurity scattering) lines.

\subsection{Lorenz number}
\begin{figure}%[H]
    \centering
	\includegraphics[scale=0.423]{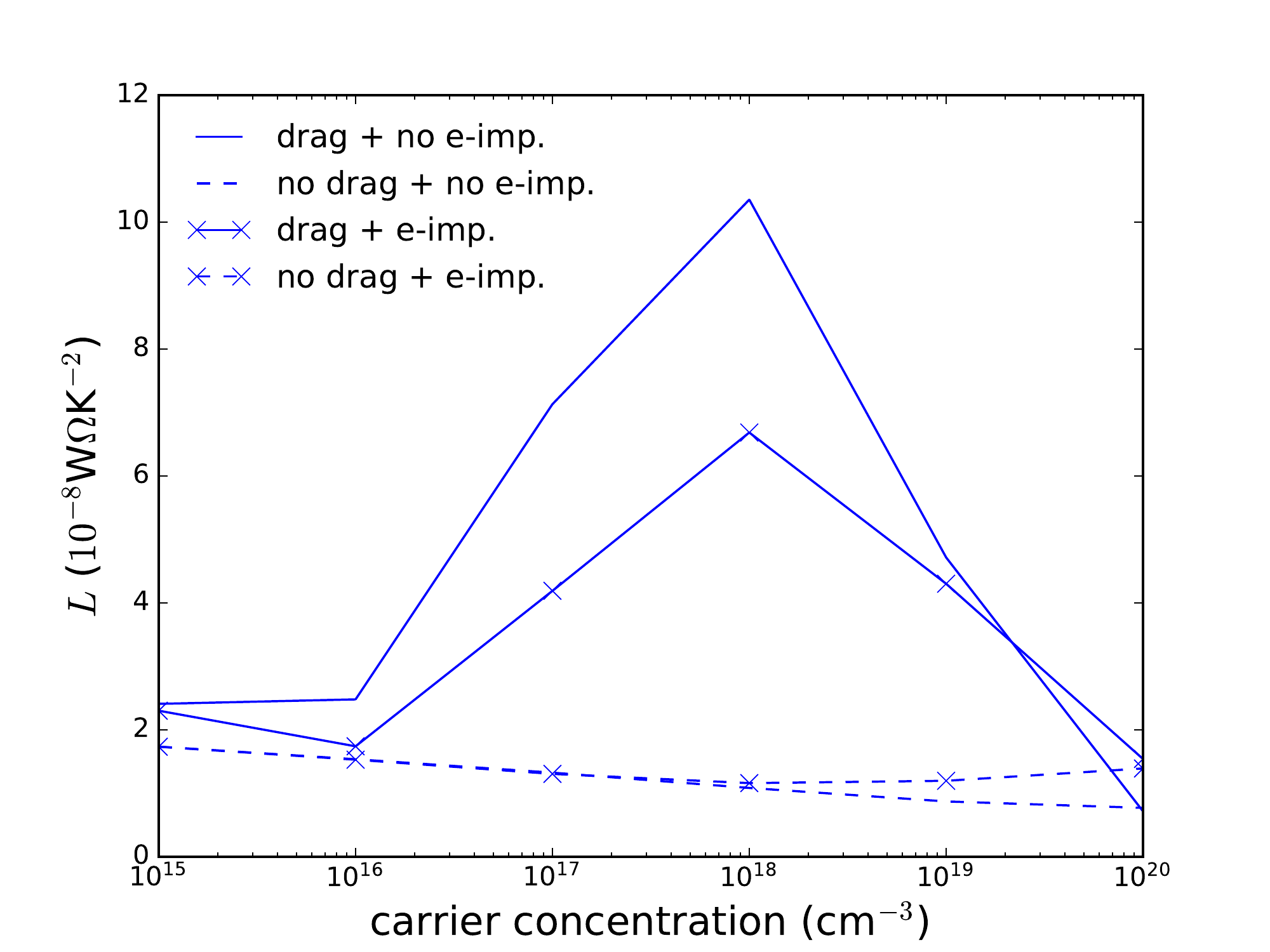}
	\caption{Lorenz number as a function of carrier concentration at $300$ K.}
	\label{fig:lorenz}
\end{figure}

Fig. \ref{fig:lorenz} shows the Lorenz number, $L$, as a function of doping concentration. For metals, the Wiedemann-Franz (WF) law value of the Lorenz number is $2.44 \times 10^{-8}$ W$\Omega$K$^{-2}$, and for semiconductors is expected to vary between $1.5$ and $2.5$ $\times 10^{-8}$ W$\Omega$K$^{-2}$ \cite{thesberg2017lorenz}. The Lorenz number is a crucial ingredient for decoupling the lattice thermal conductivity $\kappa_{\text{ph}}$ and the electronic contribution $\kappa_{\text{el}}$ from measurements of the total $\kappa$ \cite{putatunda2019lorenz}. While the deviations of $L$ from the metallic limit are expected in materials that exhibit significant inelastic scattering, our new finding is that it is the drag effect that leads to exceptionally high $L$ values in 3C-SiC over a wide range of carrier concentrations. This strong violation of the WF law is a consequence of the fact that the electron $\kappa$ has a stronger drag enhancement compared to the mobility over a large doping range.

\subsection{Thermopower}
\begin{figure}%[H]
    \centering
	\includegraphics[scale=0.423]{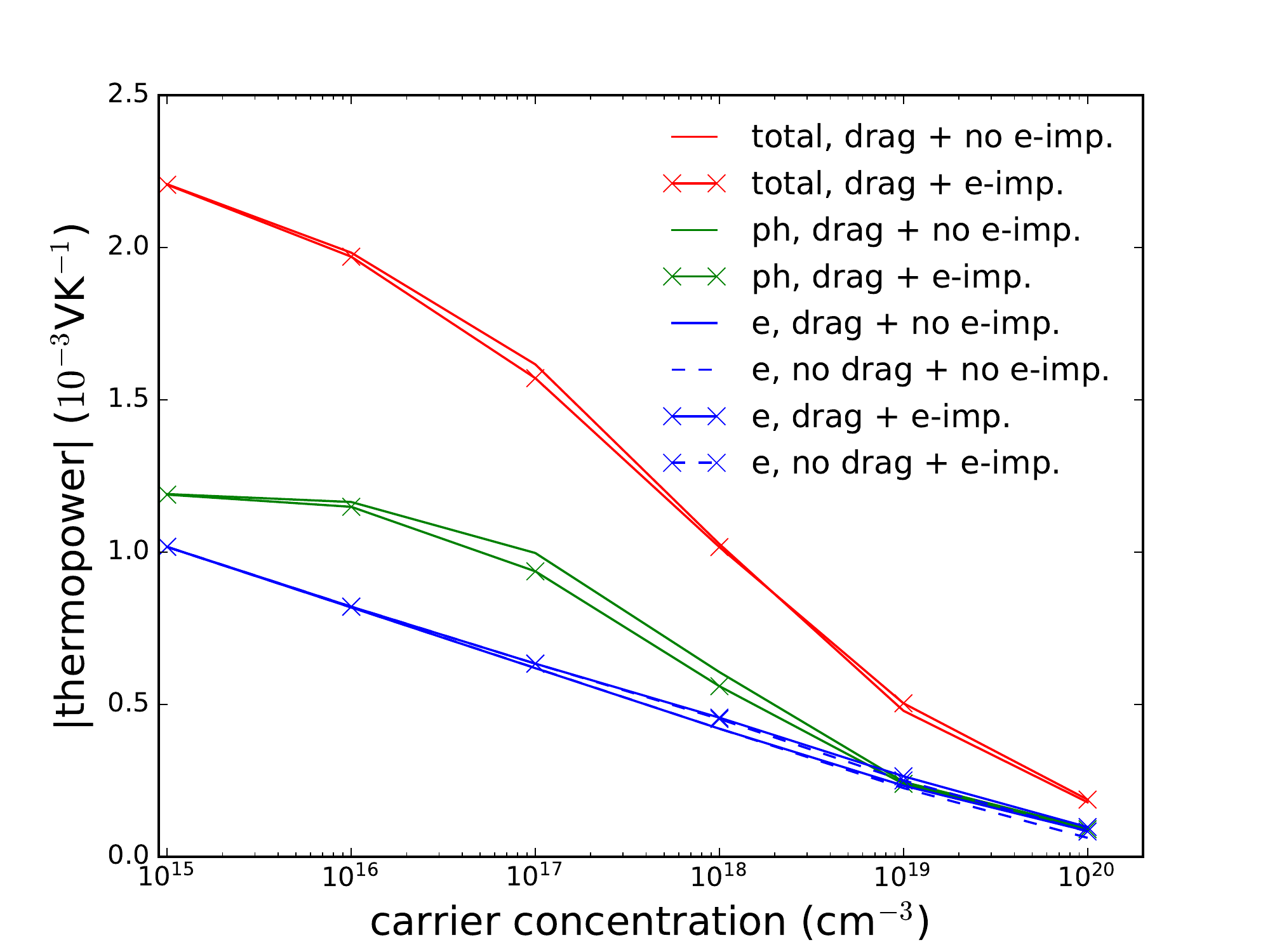}
	\caption{Absolute value of the total thermopower and its electron and phonon components as a function carrier concentration at $300$ K.}
	\label{fig:thermo}
\end{figure}
\begin{figure}%[H]
    \centering
	\includegraphics[scale=0.423]{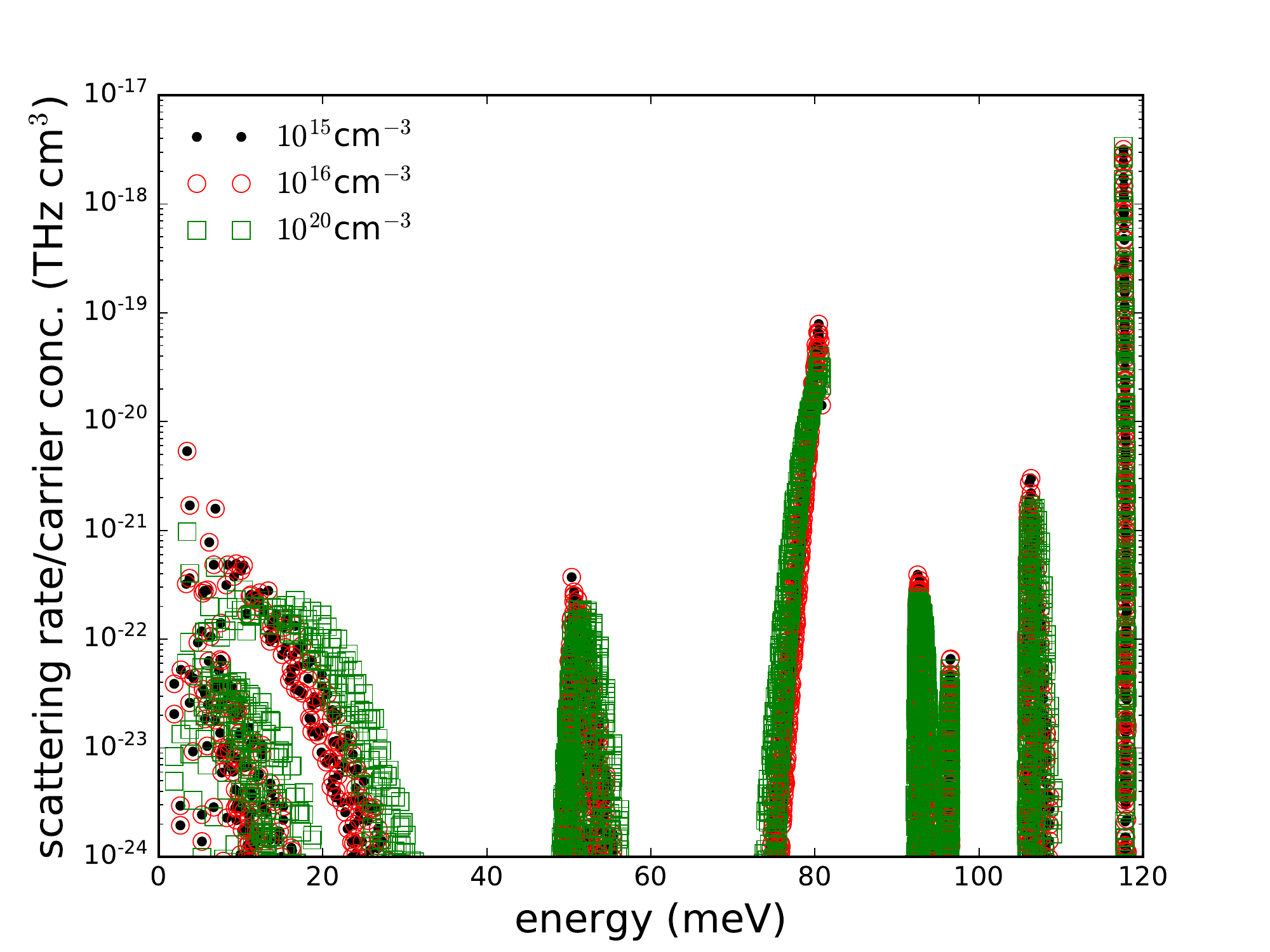}
	\caption{Phonon-electron scattering rates scaled by the carrier concentration at $300$K.}
	\label{fig:phebyconc}
\end{figure}

We now examine the absolute value of the thermopower, $|Q|$, in the Peltier picture in Fig. \ref{fig:thermo}. The Peltier picture provides a clear separation of the phonon and electron contributions, $|Q_{\text{ph}}|$ and $|Q_{\text{el}}|$, respectively, to the thermopower since both phonons and electrons can carry heat. Such a clean separation is not possible within the Seebeck picture since phonons do not carry charge. We show in the Appendix that, within numerical errors, the Peltier and the Seebeck pictures give the same thermopower, in accordance with the Kelvin-Onsager relation \cite{sondheimer1956kelvin}. $|Q_{\text{ph}}|$ is non-zero only when the phonon-electron interaction is present, since the phonon system does not explicitly couple to the applied electric field. As such, any non-zero phonon contribution is purely an electron drag effect. Surprisingly, we find that in the low doping limit $|Q_{\text{ph}}|$ in the fully coupled BTE solution is noticeably higher than $|Q_{\text{el}}|$, whereas in the high doping limit, they are nearly equal. This exceptionally strong drag effect is largely a consequence of the predominance of the scattering of electrons by small energy acoustic phonons, which pumps excess momentum into these phonons, as well as the relatively large lifetimes of these phonons, which allows them to retain the excess momentum without dissipation. As the carrier concentration decreases, $|Q_{\text{ph}}|$ is expected to approach a constant \cite{herring1954theory}. First, we consider the drag effect without the impurity channel for electron scattering. For the acoustic phonons the carrier concentration dependence comes only from the reduction of the scattering phase space of electrons at the chemical potential by the $X$-point LA phonons, while low energy acoustic phonon scattering is unaffected by the location of the chemical potential. At low carrier concentrations, when the chemical potential is in the band gap, the rate at which low energy acoustic phonons receive momentum from electrons remains nearly the same as a function of carrier concentration. Since the phonon-phonon scattering rates are independent of the carrier concentration in our rigid band model, and since the phonon-electron scattering rates scale linearly with the concentration in the low doping limit, as can be seen in Fig. \ref{fig:phebyconc}, the total amount of momentum received from the electron system that is sustained in the phonon system thus approaches a constant with decreasing carrier concentration. With increasing carrier concentration, the phonon-electron scattering rates begin to dominate the phonon-phonon scattering rates and progressively more of the excess momentum is returned to the electronic system. As a consequence, $|Q_{\text{ph}}|$ decreases with increasing carrier concentration. This has been described as the ``saturation effect" \cite{herring1954theory}. 

% In Fig. \ref{fig:phebyconc} the phonon-electron scattering rates scaled by the carrier concentration is plotted for three different carrier concentrations. In the low doping limit, the phonon-electron scattering rates are found to be nearly exactly proportional to the carrier concentration (black points and open red circles). This results in the constancy of the phonon component of the thermopower as discussed in the text. At the high concencentration limit, the phonon-electron scattering rates increase sublinearly with respect to the carrier concentration (open green squares). The consequence of this is the ``saturation effect" as discussed in the text.

\begin{figure}%[H]
    \centering
	\includegraphics[scale=0.423]{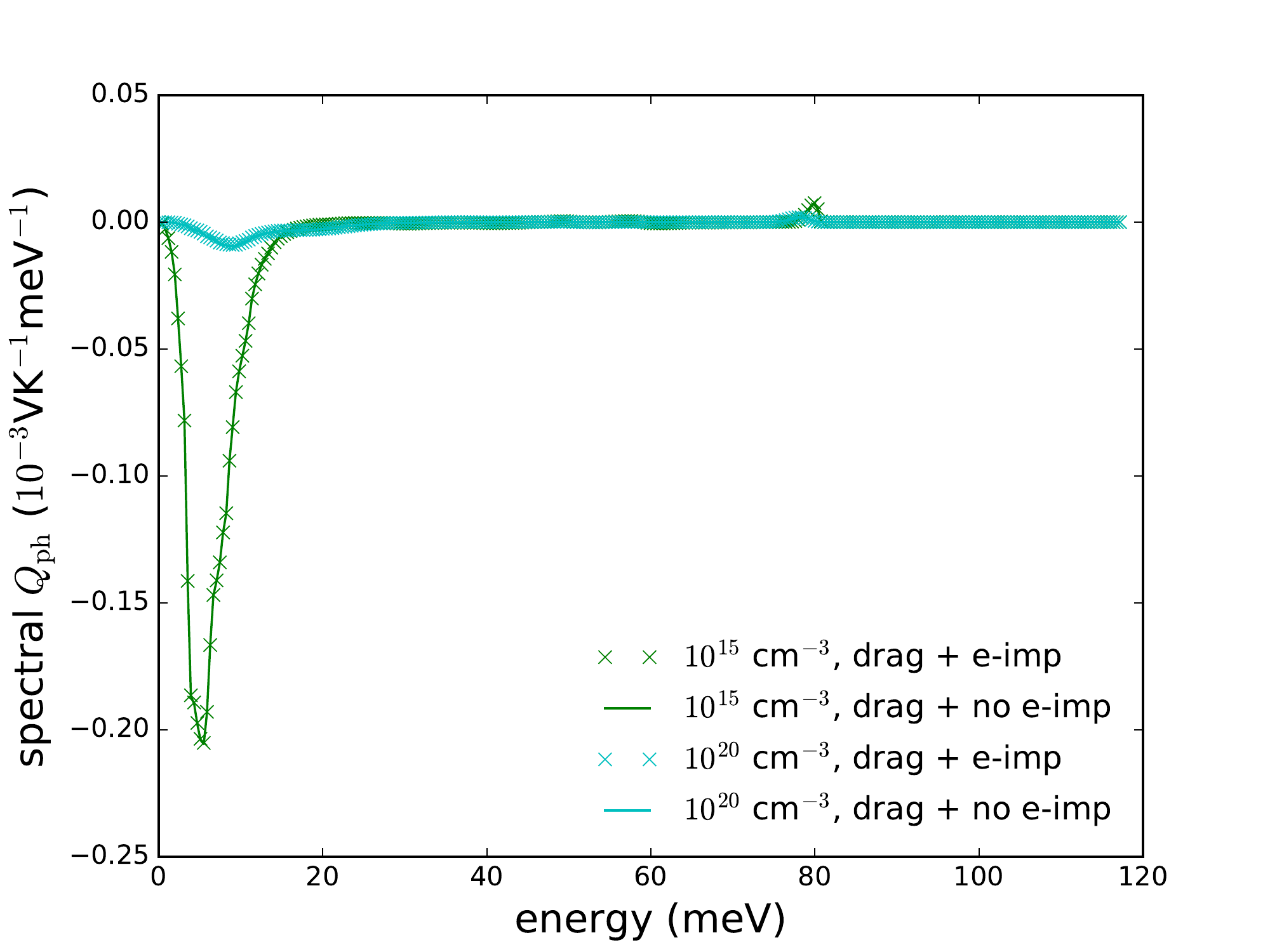}
	\caption{Spectral decomposition of the phonon contribution to thermopower $Q$ for carrier concentrations $10^{15}$ and $10^{20}$ cm$^{-3}$.}
	\label{fig:specQ_ph}
\end{figure}

This is demonstrated further in Fig. \ref{fig:specQ_ph} which shows the spectrum of the phonon contribution to the thermopower $Q$ for carrier concentrations $10^{15}$ (blues symbols) and $10^{20}$ cm$^{-3}$ (cyan symbols). In both cases, there is no effect of the electron-charged impurity scattering channel (crosses) on $Q_\text{ph}$, which we will explain shortly. Note that when the phonon BTE is decoupled from the electron BTE in the presence of an external electric field, the phonon contribution to $Q$ is trivially zero. For the low concentration case (green line), there is strong drag effect on the low energy acoustic phonons. These phonons receive momentum from the electronic system due to strong electron-phonon coupling. Since the external electric field does not couple to phonons, the non-drag activated high velocity acoustic phonons are not driven out of equilibrium and, as such, only the small energy acoustic phonons contribute to $Q$. For the high concentration case, phonon-electron scattering rates are higher than phonon-phonon ones and, as such, the phonons return the momentum received from the electrons. This leads to a reduction in the phonon contribution of $Q$ compared to the low doped case.

On the other hand, the drag enhancement of $|Q_{\text{el}}|$ is negligible. The reason for this lies in the fact that $|Q_{\text{el}}|$ is proportional to the ratio of the carrier heat and charge current densities, both of which are boosted by the phonon drag with increasing carrier concentration, leading to significant cancellation. Very similar arguments explain the strong drag effect on $|Q|$ in the Seebeck picture which we discuss below.

The Seebeck picture, where the thermopower is the response of the electronic system to an applied temperature gradient field under an open circuit condition, seemingly poses a puzzle: how can there be a strong phonon drag boost of the Seebeck $|Q|$ at low carrier concentrations where none exists for the electronic contribution to the thermopower, thermal conductivity, and mobility? We first note that the drift of phonons in a Seebeck experiment occurs due to a direct coupling to the applied temperature gradient field, and not merely as a secondary effect due to phonon-electron coupling. In the non-degenerate limit, as the carrier concentration is decreased, the phonon-electron scattering rates also diminish proportionally. Therefore, the rate of momentum transferred from the long-lived low energy acoustic phonons to the electron system is proportional to the carrier concentration. However, as there are now proportionally fewer electrons in the system, the momentum gain from the phonon system per electron is a constant. For the build-up of a Seebeck voltage, it is the momentum gain per electron that matters since this allows more electrons to overcome the growing, self-consistent electric field during the transient period. On the other hand, with increasing carrier concentration, the phonon-electron scattering rates increase sub-linearly, as shown earlier. Thus, in the high doping limit, increasing the carrier concentration decreases the momentum gain per electron, hence the phonon drag gain of $|Q|$ - the ``saturation effect" occurs. A similar analysis has been done for a partially coupled calculation in Ref. \cite{zhou2015ab}.

Lastly, we discuss the striking insensitivity of the thermopower to the presence of impurity scattering. In the Seebeck picture, the rate of momentum received by the phonons from the temperature gradient field and, thus, the momentum transfer to the electronic system remains the same as before. With increasing doping concentrations, the rate of draining of momentum from the electronic system in the impurity channel increases. As such the same steady state voltage will develop in the end. In other words, the total momentum received per electron from the phonon system remains the same regardless of the presence of impurities. Similar arguments hold in the Peltier picture in terms of the constancy of the momentum retaining capacity of the phonons in the presence of an impurity scattering channel in the electronic system. $|Q_{\text{el}}|$ is unaffected by impurity scattering for similar reasons which has previously been demonstrated by \textit{ab initio} calculations in Ref. \cite{fiorentini2016thermoelectric}. For the drag component of the thermopower the same has been shown in Ref. \cite{zhou2015ab}.  We have numerically verified this phenomenon by artificially increasing the electron-charged impurity scattering rates by a factor of $100$ at both the $10^{15}$ and $10^{20}$ cm$^{-3}$ doping concentrations and found that the same $|Q_{\text{ph}}|$ and $|Q_{\text{el}}|$ as before are reproduced. In the Appendix we present a similar analysis to show that $|Q_{\text{ph}}|$ and $|Q_{\text{el}}|$ are also largely unaffected by phonon-isotope scattering, corroborating the results in Ref. \cite{zhou2015ab}.

\section{Conclusion} \label{conclusion}
In this work we study combined non-equilibrium dynamics of electrons and phonons resulting in the mutual drag effect by solving fully coupled electron and phonon Boltzmann transport equations, for the first time treating both electron-phonon and phonon-phonon coupling at the \textit{ab initio} level and taking into account impurity scattering. In the case of 3C-SiC at room temperature, we found that the intrinsic electron mobility is significantly enhanced by the phonon drag, while the phonon thermal conductivity is weakly affected. We saw that phonon transport accounts for a remarkably large contribution to the thermopower, over a wide doping range, contrary to the conventional picture in semiconductors. Also, the electron-phonon drag causes a significant increase in the Lorenz number, outside the range previously expected in semiconductors, which affects how lattice thermal conductivity measurements should be interpreted. These are consequences of the strong piezoelectric and acoustic deformation potential type scattering of electrons by low-energy acoustic phonons and optical deformation type scattering by the zone boundary LA phonons, and the large LO phonon energy in this material. Since such features are typically absent in non-polar or weakly polar materials, we expect room temperature drag effect to be weaker in those materials compared to that in 3C-SiC. The presence of impurity scattering suppresses the strong drag enhancement of the electron mobility and thermal conductivity, while having a much weaker effect on the drag enhancement of the Lorenz number and thermopower. Based on this analysis, we expect that the hexagonal polytypes - 2H-, 4H-, and 6H-SiC will and a wide range of polar semiconductors in general will also exhibit similarly strong drag phenomena even at room temperature.

\section*{Acknowledgements}
This work was supported by the STC Center for Integrated Quantum Materials, NSF Grant No. DMR-1231319 and the US Department of Energy (DOE) Office of Basic Energy Sciences under Award No. DE-SC0020128. We also acknowledge generous support from the Harvard University Climate Change Solutions Fund.
\\
%\newpage
\appendix

\section*{Appendix}

\subsection{Coupled BTEs} 
Here we provide the expressions for the various terms in the BTEs given in Eq. 3 in the main text. We use the following notation when needed for an electronic state: $\nu \equiv (m,\mb{k})$, where $m$ is the electronic band index and $\mb{k}$ is the wave vector. Similarly, a phonon mode is denoted by $\lambda \equiv (s, \mb{q})$, where $s$ is the branch index and $\mb{q}$ is the wave vector. Furthermore, we use the notation $[\mb{k} + \mb{q}]$ to denote $(\mb{k} + \mb{q})$ modulo $\mb{G}$, where $\mb{G}$ is the reciprocal lattice vector.

The phonon absorption (+) and emission (-) transition probabilities for electrons are given by
\begin{widetext}
\begin{align}\label{eq:X}
\stirling{X^{+}_{\nu\nu'\lambda}}{X^{-}_{\nu\nu'\lambda}} = \dfrac{2\pi}{\hbar}\gsq \stirling{ \fdsub{m\kk}(1-\fdsub{n[\kk+\qq]})\besub{s\qq}\delta(\elen{n[\kk+\qq]}-\elen{m\kk}-\phen{s\qq}) } { \fdsub{m\kk}(1-\fdsub{n[\kk+\qq]})(1+\besub{s-\qq})\delta(\elen{n[\kk+\qq]}-\elen{m\kk}+\phen{s-\qq}) },
\end{align}
\end{widetext}
where $\epsilon_{m\kk}$ is the electron energy and $\hbar\omega_{s\qq}$ is the phonon energy.

The phonon-electron scattering transition probabilities are identical to the + processes given above, but for clarity we write it separately as
\begin{align}\label{eq:Y}
Y_{\lam mn\kk} = &\dfrac{2\pi}{\hbar}\gsq\\ \nonumber 
&\times \fdsub{m\kk}(1-\fdsub{n[\kk+\qq]})\besub{\lam}\delta(\elen{n[\kk+\qq]}-\elen{m\kk}-\phen{\lam}).
\end{align}

The third order phonon-phonon transition probabilities are \cite{li2014shengbte}
\begin{align}\label{eq:W}
W^{\pm}_{\lam\lam'\lam''} = &\dfrac{\pi\hbar}{4}\dfrac{|V^{\pm}_{\lam\lam'\lam''}|^{2}}{\omega_{\lam}\omega_{\lam'}\omega_{\lam''}} \\ \nonumber 
&\times (\besub{\lam}+1)\left(\besub{\lam'}+\dfrac{1}{2} \pm \dfrac{1}{2}\right)\besub{\lam''} \delta(\omega_{\lam}\pm\omega_{\lam'}-\omega_{\lam''}).
\end{align}

We can collect the total electronic out-scattering probability in
\begin{equation}\label{eq:R}
	R_{\nu} = \sum_{n\lambda} \left(X^{+}_{\nu n[\kk+\qq]\lam} + X^{-}_{\nu n[\kk+\qq]\lam}\right) + R^{\text{imp}}_{\nu},
\end{equation}
where $R^{\text{imp}}_{\nu}$ is the charged impurity scattering term.

In terms of the above, electronic RTA scattering rates are given by
\begin{equation}\label{eq:We}
    W^{\text{e,RTA}}_{\nu} = \dfrac{R_{\nu}}{\fdsub{\nu}\left(1-\fdsub{\nu}\right)}.
\end{equation}

Similarly, for the phonon system we have
\begin{equation}\label{eq:Q}
	Q_{\lam} = \sum_{\lam'\lam''}\left(W^{+}_{\lam\lam'\lam''}+\dfrac{1}{2}W^{-}_{\lam\lam'\lam''}\right) + 2\sum_{mn\kk}Y_{\lam mn\kk}  + Q^{\text{iso}}_{\lam},
\end{equation}
where the prefactor $2$ of $Y$ is due to the spin degrees of freedom and $Q^{\text{iso}}_{\nu}$ denotes the isotope scattering term.

The phonon RTA scattering rates are
\begin{equation}\label{eq:Wph}
    W^{\text{ph,RTA}}_{\lam} = \dfrac{Q_{\lam}}{\besub{\lam}(\besub{\lam}+1)}.
\end{equation}

In this study we considered two fields and four equations. We give the expression for them below.

\textbf{Electron response to temperature gradient field:}

The field coupling term is
\begin{equation}\label{eq:I0}
    \Isub{\nu}^{0} = \dfrac{\elen{\nu}-E_{\text{F}}}{R_{\nu}T}\fdsub{\nu}\left(1-\fdsub{\nu}\right)\velsub{\nu},
\end{equation}
where $E_{\text{F}}$ is the chemical potential and $\mb{v}_{\nu}$ is the electronic group velocity. This is the RTA term.

The self term is given by
\begin{equation}\label{eq:IS}
	\Delta \Isub{\text{S},\nu} = \dfrac{1}{R_{\nu}}\sum_{ns\qq} \Isub{n[\kk+\qq]}\left(X^{+}_{\nu n[\kk+\qq]\lam} + X^{-}_{\nu n[\kk+\qq]\lam}\right),
\end{equation}
and the drag term is given by
\begin{equation}\label{eq:ID}
	\Delta \Isub{\text{D},\nu} = \dfrac{1}{R_{\nu}}\sum_{ns\qq} \left(X^{-}_{\nu n[\kk+\qq]\lam}\Fsub{s-\qq} - X^{+}_{\nu n[\kk+\qq]\lam}\Fsub{s\qq}\right).
\end{equation}

\textbf{Phonon response to temperature gradient field:}

The field coupling term is given by

\begin{equation}\label{eq:F0}
	\Fsub{\lam}^{0} = \dfrac{\phen{\lam}\velsub{\lam}\besub{\lam}(\besub{\lam}+1)}{Q_{\lam}T},
\end{equation}
where $\velsub{\lam}$ is the phonon group velocity. This is the RTA term.

The self and the drag terms are given by
\begin{align}\label{eq:FS}
	\Delta\Fsub{\text{S},\lam} = \dfrac{1}{Q_{\lam}}\sum_{\lam'\lam''}\big[ &W^{+}_{\lam\lam'\lam''}\left( \Fsub{\lam''}-\Fsub{\lam'} \right) \\ \nonumber
	&+ \dfrac{1}{2}W^{-}_{\lam\lam'\lam''}\left( \Fsub{\lam''}+\Fsub{\lam'} \right) \big],
\end{align}

\begin{equation}\label{eq:FD}
	\Delta\Fsub{\text{D},\lam} = \dfrac{2}{Q_{\lam}}\sum_{mn\kk}Y_{\lam mn\kk}\left(\Isub{n[\kk+\qq]}-\Isub{m\kk}\right).
\end{equation}

\textbf{Electron response to electric field:}

The field coupling, RTA term is given by
\begin{equation}\label{eq:J0}
    \Jsub{\nu}^{0} = \dfrac{e}{R_{\nu}}\fdsub{\nu}\left(1-\fdsub{\nu}\right)\velsub{\nu}.
\end{equation}

The self and drag terms are
\begin{equation}\label{eq:JS}
    \Delta \Jsub{\text{S},\nu} = \dfrac{1}{R_{\nu}}\sum_{ns\qq} \Jsub{n[\kk+\qq]} \left(X^{+}_{\nu n[\kk+\qq]\lam} + X^{-}_{\nu n[\kk+\qq]\lam}\right),
\end{equation}

\begin{equation}\label{eq:JD}
    \Delta \Jsub{\text{D},\nu} = \dfrac{1}{R_{\nu}}\sum_{ns\qq} \left(X^{-}_{\nu n[\kk+\qq]\lam}\Gsub{s-\qq} - X^{+}_{\nu n[\kk+\qq]\lam}\Gsub{s\qq}\right).
\end{equation}

\textbf{Phonon response to electric field:}

There is no field coupling between phonons and the electric field. The self and drag terms are

\begin{align}\label{eq:GS}
\Delta\Gsub{\text{S},\lam} = \dfrac{1}{Q_{\lam}}\sum_{\lam'\lam''}\big[ &W^{+}_{\lam\lam'\lam''}\left( \Gsub{\lam''}-\Gsub{\lam'} \right) \\ \nonumber 
&+ \dfrac{1}{2}W^{-}_{\lam\lam'\lam''}\left( \Gsub{\lam''}+\Gsub{\lam'} \right) \big],
\end{align}

\begin{equation}\label{eq:GD}
\Delta\Gsub{\text{D},\lam} = \dfrac{2}{Q_{\lam}}\sum_{mn\kk}Y_{\lam mn\kk}\left(\Jsub{n[\kk+\qq]}-\Jsub{m\kk}\right).
\end{equation}

\subsection{Code validation: comparison to simple models}
To validate our code, we compared the \textit{ab initio} 3C-SiC RTA scattering rates to those from simple analytical models. We first discuss briefly the simple model calculations. In the model system we assume that the free carriers in the system form a homogeneous electron gas described by an isotropic, parabolic electron band with effective mass $m^{*}$. For zone center and zone edge acoustic phonons we consider the acoustic deformation potential (ADP) and zeroth order optic deformation potential (ODP) type scattering, respectively. The matrix elements for these processes take the form of \cite{nag2012electron}
\begin{equation}
    g^{\text{DP}}_{q} = \sqrt{\dfrac{\hbar}{2V\rho\omega_{q}}}M_{q},
\end{equation}
where $V$ is the volume and $\rho$ is the mass density with
\begin{align}
    M^{\text{ADP}}_{q} &= D^{\text{A}}q \text{ and} \nonumber \\
    M^{\text{ODP}}_{q} &= D^{\text{O}},
\end{align}
where $q$ is understood to be the magnitude of the phonon wave vector connecting two electronic states. $D^{\text{A}}$ and $D^{\text{O}}$ are the acoustic and optic deformation potentials.

We assume that the near zone center LA phonons take part in ADP scattering, whereas the LA phonons near the $X$-point take part in ODP scattering. For the latter the degeneracy effect is taken into account by multiplying the scattering rates with an appropriate degeneracy factor \cite{sanborn1992theory}.

Since 3C-SiC is a non-centrosymmetric, strongly polar material, the piezoelectric scattering channel is also considered. As is customary, we combine all acoustic branches into one effective piezo-active branch. This is done by averaging the zone center TA and LA branch speeds to obtain an effective speed $v_{\text{PZ}}$. The matrix element for this process is \cite{nag2012electron}
\begin{equation}
    g^{\text{PZ}}_{q} = \sqrt{\dfrac{\hbar e^{2} e^{2}_{\text{PZ}}}{2V\rho v_{\text{PZ}}\epsilon^{2}_{0}\kappa^{2}_{\infty}q} }\left(\dfrac{q^{2}}{q^{2}+q^{2}_{\text{TF}}}\right),
\end{equation}
where $e_{\text{PZ}}$ is the piezoelectric scattering strength, $\epsilon_{0}$ is the permittivity of free space, $\kappa_{\infty}$ is the high-frequency dielectric constant, and $q_{\text{TF}}$ is the Thomas-Fermi screening wave vector computed assuming a homogeneous electron gas. The term in the brackets is due to Thomas-Fermi screening, and is required for this scattering mechanism to regularize the singularity at the conduction band edge in the corresponding scattering rates expression which we will present shortly. 

Lastly, we considered the polar optic phonon (POP) scattering from the zone center LO phonons using the Fr\"{o}hlich interaction \cite{nag2012electron}
\begin{equation}
    g_{q}^{\text{POP}} = \sqrt{\dfrac{\hbar e^{2} \omega_{\text{LO}}}{2V\epsilon_{0}q^{2}} \left(\dfrac{1}{\kappa_{\infty}} - \dfrac{1}{\kappa_{0}}\right)}, 
\end{equation}
where $\omega_{\text{LO}}$ is the angular frequency of the LO phonon at the zone center and $\kappa_{0}$ is the static dielectric constant.

Obtaining the RTA scattering rates from these matrix elements involve performing elementary integrals of the electron-phonon collision term. For two dimensional systems, such calculations are demonstrated in Ref. \cite{kaasbjerg2012phonon}. The generalization to three dimensions is straightforward and here we simply provide the final scattering rates expressions. For the zone-center ADP channel we get
\begin{equation}
    W^{\text{ADP}}(E_{k}) = 
    \dfrac{2^{3/2}(D^{A})^{2}k_{\text{B}}T (m^{*})^{3/2}\sqrt{E_{k}}}{2\pi\hbar^{4}\rho v_{\text{LA}}^{2}},
\end{equation}
where $m^{*}$ is the effective mass density of states mass, $E_{k}$ is the electron energy measured from the conduction band minimum (CBM), and $v_{\text{LA}}$ is the LA phonon speed.

The $X$-point LA phonon ODP scattering rates can be shown to be
\begin{align}
    W^{\text{ODP}}(E_{k}) &= \dfrac{(g_{\text{V}}-1)(m^{*})^{3/2}(D^{O})^{2}}{\sqrt{2}\pi\hbar^{3}\rho\omega_{X,\text{LA}}\left[1 - f^{0}(E_{k})\right]} n^{0}(\hbar\omega_{X,\text{LA}}) \nonumber \\
    \times \Bigg\{ & \big[1 - f^{0}(E_{k} + \hbar\omega_{X,\text{LA}}) \big] \sqrt{E_{k} + \hbar\omega_{X,\text{LA}}} \\ \nonumber
    &+ \big[1 - f^{0}(E_{k} - \hbar\omega_{X,\text{LA}}) \big] \sqrt{E_{k} - \hbar\omega_{X,\text{LA}}} \\ \nonumber 
    &\times \text{e}^{\frac{\hbar\omega_{X,\text{LA}}}{k_{\text{B}T}}}\Theta(E_{k} - \hbar\omega_{X,\text{LA}}) \Bigg\},
\end{align}
where $\omega_{X,\text{LA}}$ is the $X$-point LA phonon angular frequency, $n^{0}$ is the Bose-Einstein distribution, and the $\Theta$ is the Heaviside Theta function. Since this is an intervalley scattering process, we multiplied the expression by $(g_{\text{V}}-1)$ where $g_{\text{V}}$ is the valley degeneracy factor following Ref. \cite{sanborn1992theory}.

The piezoelectric scattering rates are given by
\begin{align}
    W^{\text{PZ}}(E_{k}) = &\dfrac{ e^{2}_{\text{PZ}}e^{2} k_{\text{B}}T m^{*} }{ 2\pi\hbar^{3}\rho v^{2}_{\text{PZ}} \kappa^{2}_{\infty} \epsilon_{0}^{2} k } \\ \nonumber 
    &\times \left[ 1 + \dfrac{1}{1 + 4k^{2}/q^{2}_{\text{TF}}} - \dfrac{q^{2}_{\text{TF}}}{2k^{2}}\text{log}\Bigg(1 + \dfrac{4k^{2}}{q^{2}_{\text{TF}}} \Bigg) \right].
\end{align}

Lastly, the polar optic phonon RTA scattering rates are
\begin{align}
    W^{\text{POP}}(E_{k}) &= \dfrac{e^{2}}{4\pi\epsilon_{0}} \left(\dfrac{1}{\kappa_{\infty}} - \dfrac{1}{\kappa_{0}}\right) \dfrac{k}{2\hbar}\dfrac{\hbar\omega_{\text{LO}}}{E_{k}}\dfrac{n^{0}(\hbar\omega_{\text{LO}})}{1 - f^{0}(E_{k})} \nonumber \\
    &\times \Bigg\{ \Big[1 - f^{0}(E_{k} + \hbar\omega_{\text{LO}})\Big]\text{log}\left| \dfrac{k+k_{+}}{k-k_{+}} \right| \\ \nonumber 
    &+\Big[1 - f^{0}(E_{k} - \hbar\omega_{\text{LO}})\Big] \text{log}\left| \dfrac{k+k_{-}}{k-k_{-}} \right| \\ \nonumber &\times\text{e}^{\frac{\hbar\omega_{\text{LO}}}{k_{\text{B}}T}} \Theta(E_{k} - \hbar\omega_{\text{LO}}) \Bigg\},
\end{align}
where $k_{\pm} \equiv \sqrt{k^{2} \pm 2m^{*}\omega_{\text{LO}}/\hbar}$.

The materials parameters are presented in Tab. \ref{tab:params}. The deformation potentials and piezoelectric strength are chosen to roughly match the results obtained from \textit{ab initio} calculations.

\begin{table}%[H]
\centering
\begin{tabular}{lcc}
\midrule
\midrule
Parameter & Symbol & Value \\
\midrule
Effective mass & $m^{*}$ & $0.35m_{\text{e}}$ kg \\
Valley degeneracy & $g_{\text{V}}$ & $3$ \\
Density & $\rho$ & $3166$ kgm$^{-3}$ \\
LO phonon energy & $\hbar\omega_{\text{LO}}$ & $118$ meV \\
$X$-point LA phonon energy & $\hbar\omega_{X,\text{LA}}$ & 80 meV \\
LA phonon speed & $v_{\text{LA}}$ & $12.5$ kms$^{-1}$ \\
Piezoelectric phonon speed & $v_{\text{PZ}}$ & $17.6$ kms$^{-1}$ \\
Acoustic deformation potential & $D^{\text{A}}$ & $10$ eV \\
Optic deformation potential & $D^{\text{O}}$ & $1.25\times 10^{11}$ eVm$^{-1}$ \\
Piezoelectric strength & $e_{\text{PZ}}$ & 0.76 Cm$^{-2}$ \\
High-frequency dielectric constant & $\kappa_{\infty}$ & $6.52$ \\
Static dielectric constant & $\kappa_{0}$ & $9.59$ \\
\midrule
\midrule
\end{tabular}
\caption{The materials parameters of 3C-SiC for simple model calculations. Above, $m_{\text{e}}$ stands for electron mass.  The deformation potentials and the piezoelectric interaction strength are from fitting the RTA scattering rates from the \textit{ab initio} calculations. The rest of the parameters are taken or calculated from Refs. \cite{sicioffe} and  \cite{sicmatproj}.} 
\label{tab:params}
\end{table}

\begin{figure}%[H]
    \centering
	\subfloat{\includegraphics[scale=0.423]{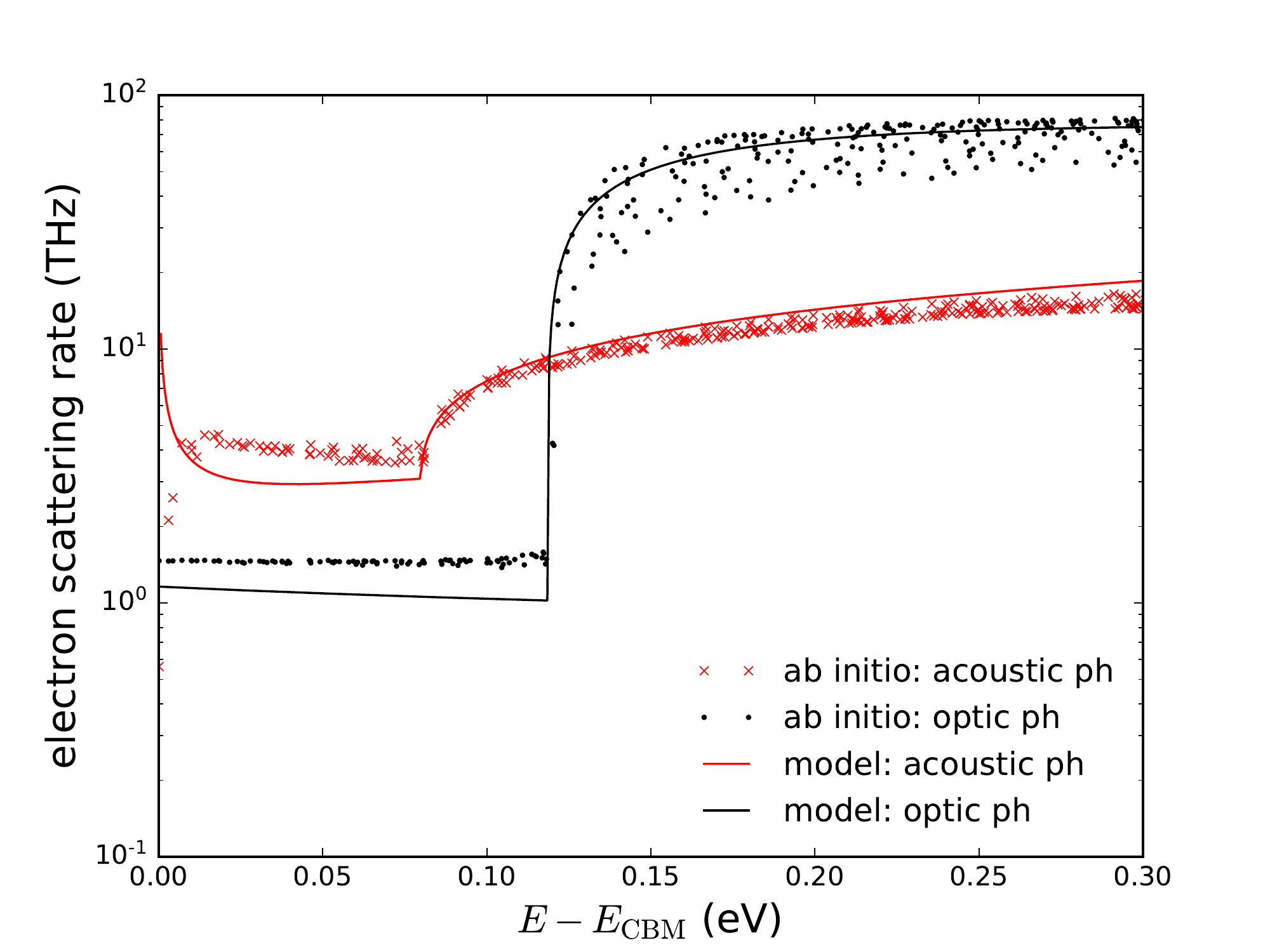}}\\
	\subfloat{\includegraphics[scale=0.423]{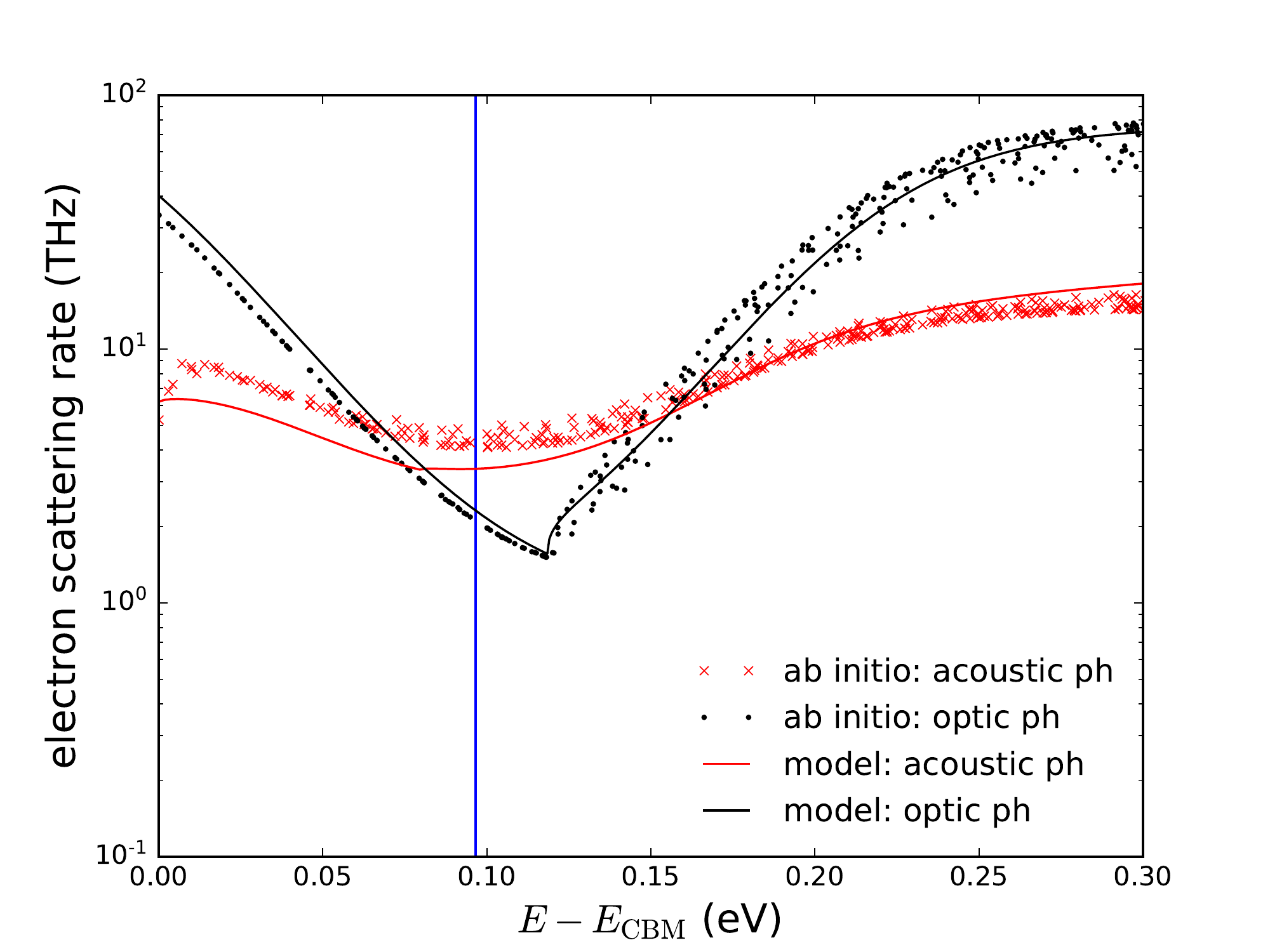}}	
	\caption{Comparison of \textit{ab initio} and simple model electronic RTA scattering rates for carrier concentrations $10^{16}$cm$^{-3}$ (top panel) and $10^{20}$cm$^{-3}$ (bottom panel) at $300$ K. The zero of the energy axis is at the conduction band minimum. For the high doping case the electron chemical potential is in the conduction band and is shown by the blue vertical line.}
	\label{fig:modeleph}
\end{figure}

In Fig. \ref{fig:modeleph} we compare the \textit{ab initio} RTA scattering rates against those from the simple model calculations. We first note that the analytic POP RTA scattering rates contain no free parameters and, as such, provide a strong check for the \textit{ab initio} calculations. We note that the \textit{ab initio} and model optic phonon scattering rates are in excellent agreement. Both show the salient physical features of the system: (1) the onset of LO phonon emission around $120$ meV (top panel) and (2) the phase-space reduction effect at the chemical potential for the high concentration case (bottom panel). Similarly, the $X$-point LA ODP model phonon scattering rates also show the emission onset around $80$ meV for the low concentration case and a noticeable phase-space reduction effect near the chemical potential for the high concentration case. The main difference between the \textit{ab initio} and the model calculation comes from the low energy acoustic phonon scattering. As mentioned before, the piezoelectric scattering rates diverge at the conduction band minimum and screening is used to regularized the divergence in the model calculation. In the \textit{ab initio} calculation we do not use screening. Instead, a small phonon energy cutoff is used, below which the scattering matrix elements are set to zero. This is a standard technique that is also used in the \texttt{EPW} software, which our code interfaces with. The low energy discrepancy between the simple model piezoelectric scattering and the \textit{ab initio} polar acoustic scattering has also been noted in Ref. \cite{liu2017first} for $n$-doped GaAs.

\subsection{Code validation: cubic Si}
\begin{figure}%[H]
    \centering
	\includegraphics[scale=0.423]{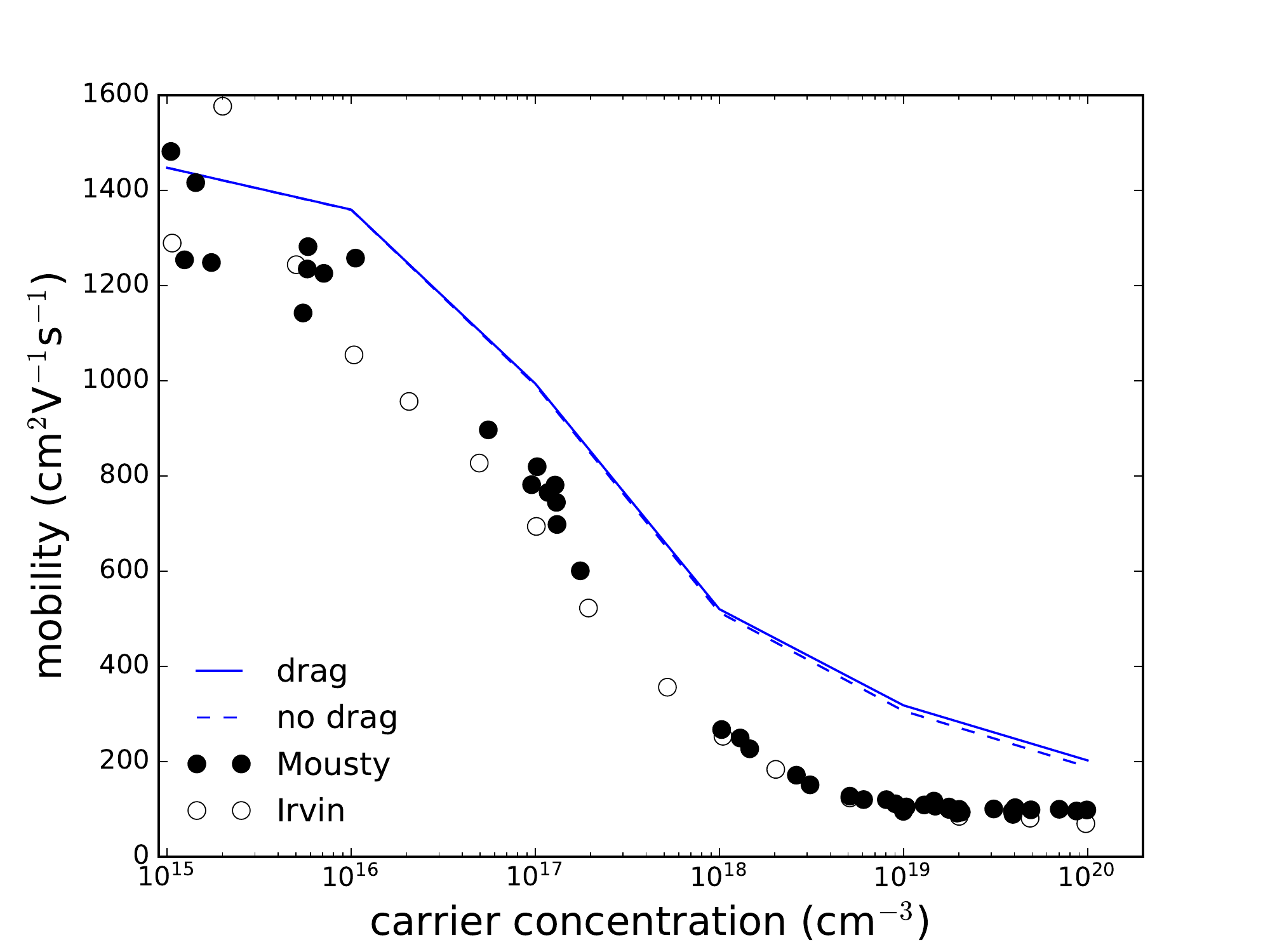}
	\caption{Mobility in silicon at $300$ K. The experimental data are taken from Ref. \cite{jacoboni1977review}.}
	\label{fig:si_mob}
\end{figure}

\begin{figure}%[H]
    \centering
	\includegraphics[scale=0.423]{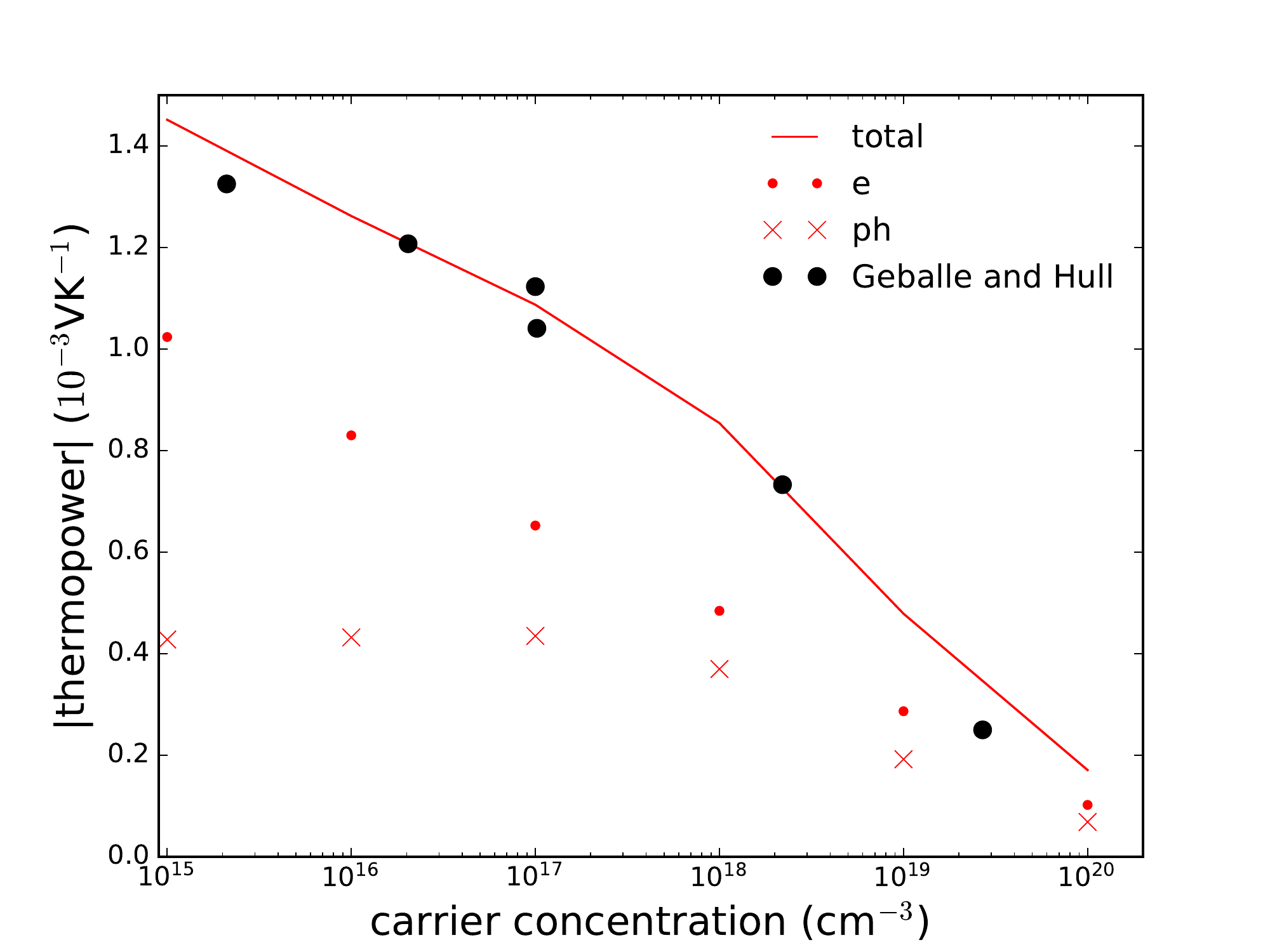}
	\caption{Thermopower breakdown in silicon at $300$ K. The experimental data is from Ref. \cite{geballe1955seebeck}.}
	\label{fig:si_thermo}
\end{figure}

To validate our code we ran it on silicon, which has previously been studied using the \textit{ab initio} partially coupled method in Refs. \cite{fiorentini2016thermoelectric} and \cite{zhou2015ab}. In Figs. \ref{fig:si_mob} and  \ref{fig:si_thermo} we present the results for a modest $(30,90)$ mesh which already gives decent agreement with measurements and previous \textit{ab initio} calculations. In particular, the moderately strong drag effect on the thermopower is accurately captured.


\subsection{Convergence}
\begin{figure}%[H]
    \centering
	\includegraphics[scale=0.423]{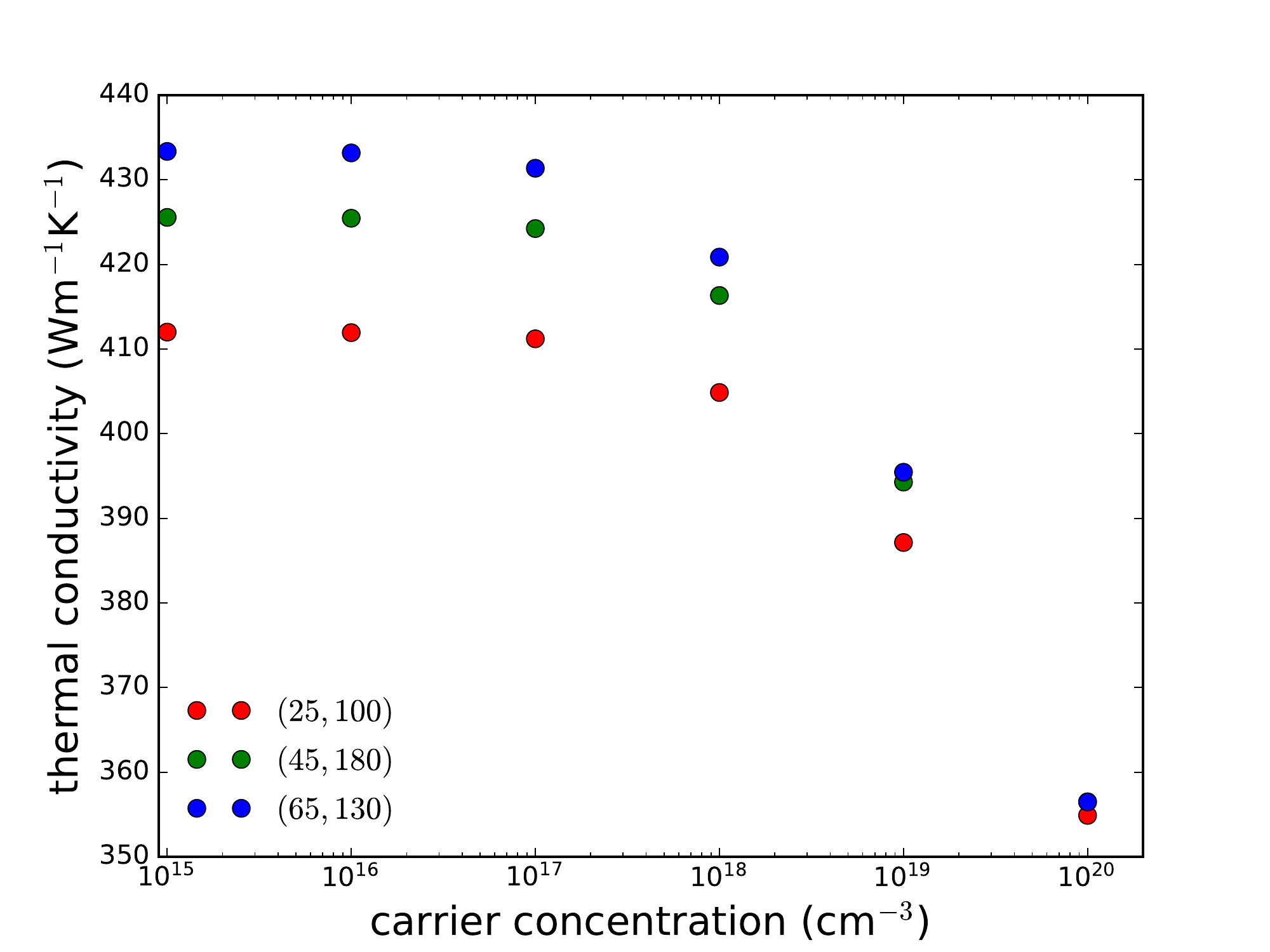}
	\caption{Phonon thermal conductivity vs. carrier concentration for various meshes.}
	\label{fig:phkappa}
\end{figure}

\begin{figure}%[H]
    \centering
	\includegraphics[scale=0.423]{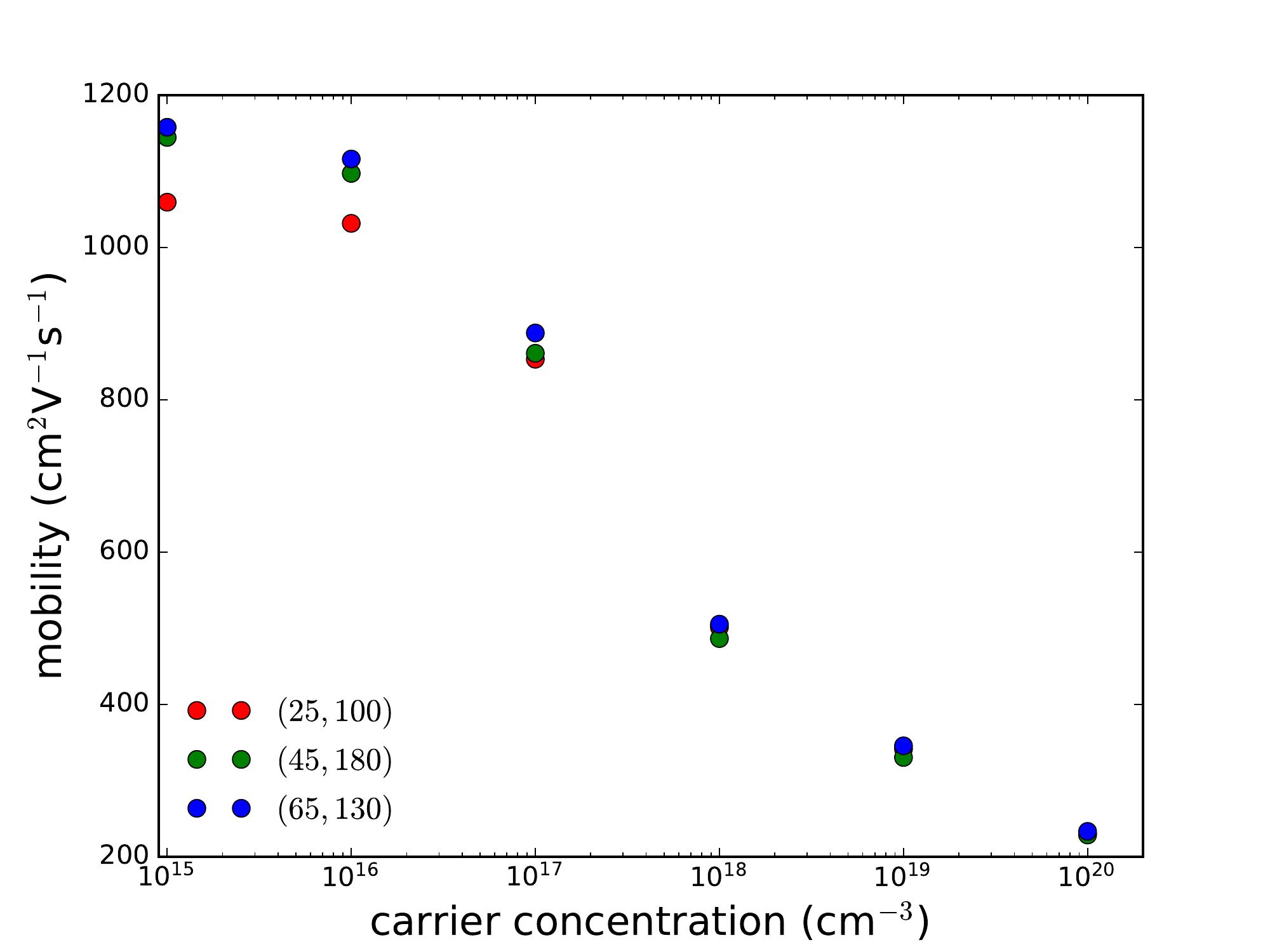}
	\caption{Carrier mobility vs. carrier concentration for various meshes.}
	\label{fig:mob}
\end{figure}

\begin{figure}%[H]
    \centering
	\includegraphics[scale=0.423]{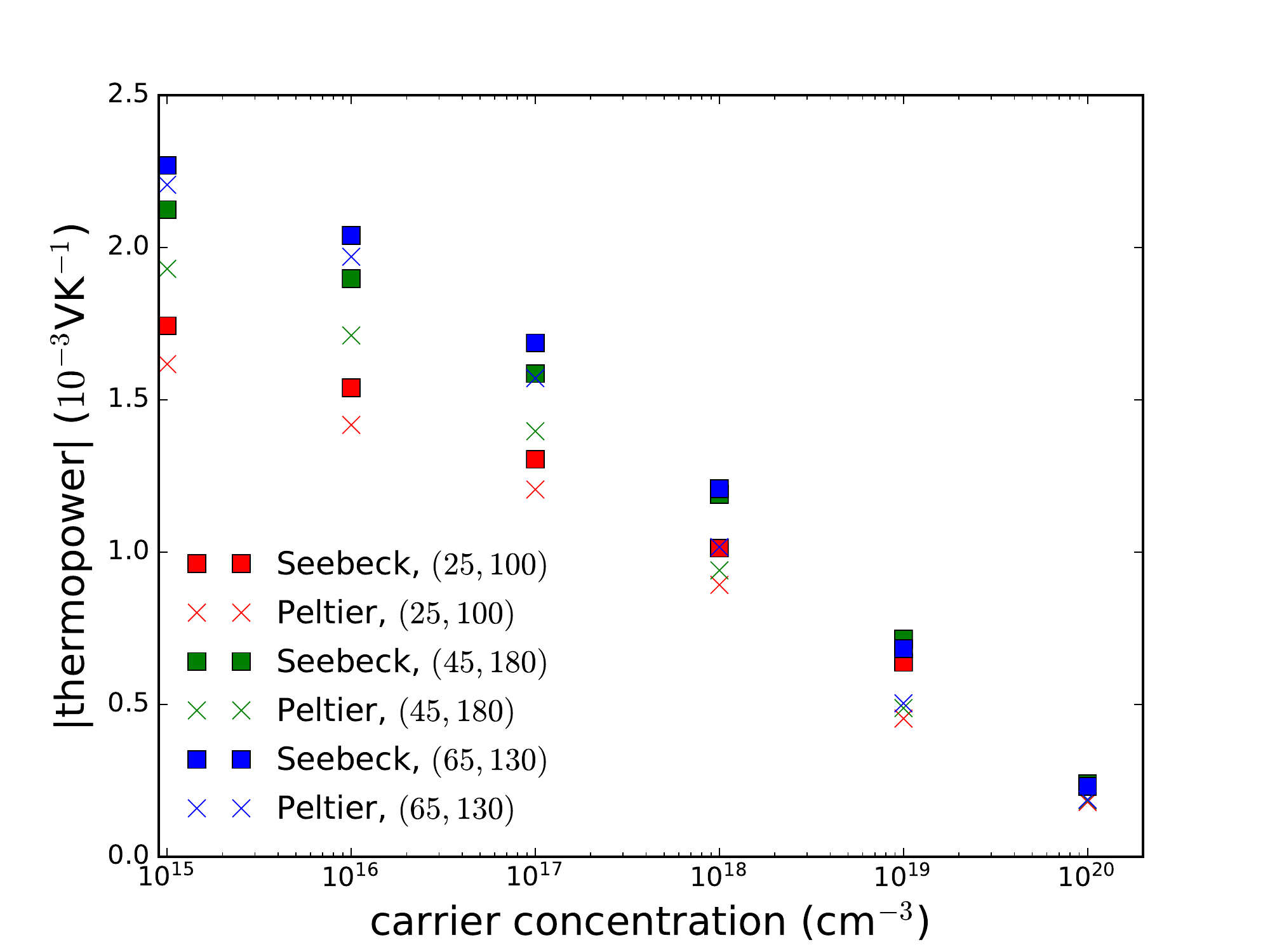}
	\caption{Thermopower in the Seebeck and Peltier pictures vs. carrier concentration for various meshes.}
	\label{fig:thermopower}
\end{figure}

Below we present the convergence test for the transport coefficients with respect to different mesh densities. For all these tests, we compare the results of the full solution of the coupled BTEs in the presence of charged impurities.
In Figs. \ref{fig:phkappa}, \ref{fig:mob}, and \ref{fig:thermopower} we present the phonon thermal conductivity, carrier mobility, and the thermopower in the Seebeck and the Peltier pictures for different mesh densities. The values obtained between the $(45, 180)$ and the $(65, 130)$ meshes are close for most of the values in the full concentration range. The Kelvin-Onsager relationship dictates that the thermopower in both the Seebeck and the Peltier pictures must match \cite{sondheimer1956kelvin}. Although we found perfect agreement at the relaxation time approximation level, in the iterated solution to the coupled BTEs the agreement deteriorates. The iteration process introduces some error due to the fact that a relatively coarse phonon mesh is used and certain coarser mesh quantities are interpolated onto the finer mesh. Moreover, the thermopower is a ratio of two different transport coefficients, and it is difficult to get the same level of accuracy for both the numerator and the denominator using the same set of meshes. Nevertheless, we see good agreement of the Kelvin-Onsager relation at the low and the high doping limits, with the largest deviation coming from the mid concentration ranges.

\subsection{Effect of phonon-isotope scattering on drag}
In Tab. \ref{tab:isodrag} we compare various transport coefficients and their drag enhancement at concentrations $10^{15}$ and $10^{20}$ cm$^{-3}$ for three cases: (1) without phonon-isotope scattering, (2) with phonon-isotope scattering, and (3) with artificially $100$ times enhanced phonon-isotope scattering. The electron-charged impurity scattering channel has been turned off. We find that the electron mobility shows negligible change at the low carrier concentration for all three cases. At the high carrier concentration, the drag enhancement decreases as the dissipative phonon-isotope scattering gets progressively stronger. For case (3), the phonon-isotope scattering rates are still small compared to the phonon-electron scattering rates for the low energy acoustic phonons, but become comparable for the optical phonons. Thus, the optic phonon drag gain to the electron mobility is reduced in this case.

Next we consider the phonon thermopower $Q_{\text{ph}}$. This quantity is not strongly affected by the presence of the dissipative isotope scattering channel. The reason for this is that the spectral contribution to $Q_{\text{ph}}$ comes from the small energy acoustic phonons and even for the artificially enhanced isotope scattering case (3), the phonon-isotope scattering rates are still significantly smaller than the phonon-electron rates. The electron thermopower $Q_{\text{el}}$ is also weakly affected by the phonon-isotope scattering channel. This is again related to the fact that the presence of a strong phonon-isotope scattering rates will only reduce the drag activity of the optic phonons while leaving the low energy acoustic phonons unaffected.

Lastly, the phonon thermal conductivity $\kappa_{\text{ph}}$ is strongly affected by the magnitude of the isotope scattering rates. The low concentration limit of $\kappa_{\text{ph}}$ reduces from $490$ to $132$ Wm$^{-1}$K$^{-1}$ between case (1) and case (3). For the high concentration case, these number are $437$ and $91$ Wm$^{-1}$K$^{-1}$. In the main text it was explained why the electron drag enhancement of $\kappa_{\text{ph}}$ is small at low carrier concentrations. The same argument applies here. For the high concentration case, the drag enhancement of $\kappa_{\text{ph}}$ increases between cases (1) and case (3). This is understood by noticing that for the low energy acoustic phonons, the isotope scattering rates are negligible even for case (3). Thus, the reduction of the $\kappa_{\text{ph}}$ due to isotope scattering comes from the mid-energy range. The drag enhancement, however, comes from the low energy acoustic phonons. Thus, the drag enhancement constitutes a progressively larger share of $\kappa_{\text{ph}}$ as we sweep from case (1) to case (3).

\begin{widetext}
\begin{table}[H]
\centering \footnotesize 
\begin{tabular}{c|c|c|c|c|c|c}
 & \multicolumn{2}{c|}{case (1) no ph-iso}  & \multicolumn{2}{c|}{case (2) ph-iso}   & \multicolumn{2}{c}{case (3) $100\times$ph-iso} \\
\midrule
conc. (cm$^{-3}$) & $10^{15}$ & $10^{20}$ & $10^{15}$ & $10^{20}$ & $10^{15}$ & $10^{20}$\\
\midrule
$\mu$ (cm$^{2}$V$^{-1}$s$^{-1}$) & 1163.51 (0.03) & 2424.66 (196.77) & 1163.50 (0.03) & 2385.49 (191.98) & 1163.40 (0.02) & 1516.22 (85.58)\\
$Q_{\text{ph}}$ ($10^{-3}$VK$^{-1}$) & -1.20 ($\infty$) & -0.10 ($\infty$) & -1.19 ($\infty$) & -0.10 ($\infty$) & -0.99 ($\infty$) & -0.07 ($\infty$)\\
$Q_{\text{el}}$ (10$^{-3}$VK$^{-1}$) & -1.02 (0.00) & -0.09 (36.27) & -1.02 (0.00) & -0.09 (35.82) & -1.02 (0.00) & -0.08 (22.80)\\
$\kappa_{\text{ph}}$ (Wm$^{-1}$K$^{-1}$) & 490.05 (0.00) & 436.68 (6.62) & 433.36 (0.00) & 380.57 (7.59) & 131.55 (0.01) & 90.74 (13.61)\\
\end{tabular}
\caption{The effect of phonon-isotope scattering on the drag enhancement of various transport coefficients in the absence of charged impurity scattering of electrons. We compare three cases: no isotope scattering, usual isotope scattering, artificially $100$ times enhanced isotope scattering. The values in the brackets are the percentage increase of the transport coefficient with respect to the corresponding non-drag value. The $\infty$ symbol implies that $Q_{\text{ph}}$ is trivially zero for the non-drag case.} 
\label {tab:isodrag}
\end{table}
\end{widetext}

%\bibliography{refs}% Produces the bibliography via BibTeX.

\begin{thebibliography}{43}%
	\makeatletter
	\providecommand \@ifxundefined [1]{%
		\@ifx{#1\undefined}
	}%
	\providecommand \@ifnum [1]{%
		\ifnum #1\expandafter \@firstoftwo
		\else \expandafter \@secondoftwo
		\fi
	}%
	\providecommand \@ifx [1]{%
		\ifx #1\expandafter \@firstoftwo
		\else \expandafter \@secondoftwo
		\fi
	}%
	\providecommand \natexlab [1]{#1}%
	\providecommand \enquote  [1]{``#1''}%
	\providecommand \bibnamefont  [1]{#1}%
	\providecommand \bibfnamefont [1]{#1}%
	\providecommand \citenamefont [1]{#1}%
	\providecommand \href@noop [0]{\@secondoftwo}%
	\providecommand \href [0]{\begingroup \@sanitize@url \@href}%
	\providecommand \@href[1]{\@@startlink{#1}\@@href}%
	\providecommand \@@href[1]{\endgroup#1\@@endlink}%
	\providecommand \@sanitize@url [0]{\catcode `\\12\catcode `\$12\catcode
		`\&12\catcode `\#12\catcode `\^12\catcode `\_12\catcode `\%12\relax}%
	\providecommand \@@startlink[1]{}%
	\providecommand \@@endlink[0]{}%
	\providecommand \url  [0]{\begingroup\@sanitize@url \@url }%
	\providecommand \@url [1]{\endgroup\@href {#1}{\urlprefix }}%
	\providecommand \urlprefix  [0]{URL }%
	\providecommand \Eprint [0]{\href }%
	\providecommand \doibase [0]{https://doi.org/}%
	\providecommand \selectlanguage [0]{\@gobble}%
	\providecommand \bibinfo  [0]{\@secondoftwo}%
	\providecommand \bibfield  [0]{\@secondoftwo}%
	\providecommand \translation [1]{[#1]}%
	\providecommand \BibitemOpen [0]{}%
	\providecommand \bibitemStop [0]{}%
	\providecommand \bibitemNoStop [0]{.\EOS\space}%
	\providecommand \EOS [0]{\spacefactor3000\relax}%
	\providecommand \BibitemShut  [1]{\csname bibitem#1\endcsname}%
	\let\auto@bib@innerbib\@empty
	%</preamble>
	\bibitem [{\citenamefont {Bloch}(1930)}]{bloch1930elektrischen}%
	\BibitemOpen
	\bibfield  {author} {\bibinfo {author} {\bibfnamefont {F.}~\bibnamefont
			{Bloch}},\ }\bibfield  {title} {\bibinfo {title} {Zum elektrischen
			widerstandsgesetz bei tiefen temperaturen},\ }\href@noop {} {\bibfield
		{journal} {\bibinfo  {journal} {Zeitschrift f{\"u}r Physik}\ }\textbf
		{\bibinfo {volume} {59}},\ \bibinfo {pages} {208} (\bibinfo {year}
		{1930})}\BibitemShut {NoStop}%
	\bibitem [{\citenamefont {Peierls}(1930)}]{peierls1930theorie}%
	\BibitemOpen
	\bibfield  {author} {\bibinfo {author} {\bibfnamefont {R.}~\bibnamefont
			{Peierls}},\ }\bibfield  {title} {\bibinfo {title} {Zur theorie der
			elektrischen und thermischen leitf{\"a}higkeit von metallen},\ }\href@noop {}
	{\bibfield  {journal} {\bibinfo  {journal} {Annalen der Physik}\ }\textbf
		{\bibinfo {volume} {396}},\ \bibinfo {pages} {121} (\bibinfo {year}
		{1930})}\BibitemShut {NoStop}%
	\bibitem [{\citenamefont {Gurevich}\ and\ \citenamefont
		{Mashkevich}(1989)}]{gurevich1989electron}%
	\BibitemOpen
	\bibfield  {author} {\bibinfo {author} {\bibfnamefont {Y.~G.}\ \bibnamefont
			{Gurevich}}\ and\ \bibinfo {author} {\bibfnamefont {O.}~\bibnamefont
			{Mashkevich}},\ }\bibfield  {title} {\bibinfo {title} {The electron-phonon
			drag and transport phenomena in semiconductors},\ }\href@noop {} {\bibfield
		{journal} {\bibinfo  {journal} {Physics Reports}\ }\textbf {\bibinfo {volume}
			{181}},\ \bibinfo {pages} {327} (\bibinfo {year} {1989})}\BibitemShut
	{NoStop}%
	\bibitem [{\citenamefont {Frederikse}\ and\ \citenamefont
		{Mielczarek}(1955)}]{frederikse1955thermoelectric}%
	\BibitemOpen
	\bibfield  {author} {\bibinfo {author} {\bibfnamefont {H.}~\bibnamefont
			{Frederikse}}\ and\ \bibinfo {author} {\bibfnamefont {E.~V.}\ \bibnamefont
			{Mielczarek}},\ }\bibfield  {title} {\bibinfo {title} {Thermoelectric power
			of indium antimonide},\ }\href@noop {} {\bibfield  {journal} {\bibinfo
			{journal} {Physical Review}\ }\textbf {\bibinfo {volume} {99}},\ \bibinfo
		{pages} {1889} (\bibinfo {year} {1955})}\BibitemShut {NoStop}%
	\bibitem [{\citenamefont {Geballe}\ and\ \citenamefont
		{Hull}(1954)}]{geballe1954seebeck}%
	\BibitemOpen
	\bibfield  {author} {\bibinfo {author} {\bibfnamefont {T.}~\bibnamefont
			{Geballe}}\ and\ \bibinfo {author} {\bibfnamefont {G.}~\bibnamefont {Hull}},\
	}\bibfield  {title} {\bibinfo {title} {Seebeck effect in germanium},\
	}\href@noop {} {\bibfield  {journal} {\bibinfo  {journal} {Physical Review}\
		}\textbf {\bibinfo {volume} {94}},\ \bibinfo {pages} {1134} (\bibinfo {year}
		{1954})}\BibitemShut {NoStop}%
	\bibitem [{\citenamefont {Geballe}\ and\ \citenamefont
		{Hull}(1955)}]{geballe1955seebeck}%
	\BibitemOpen
	\bibfield  {author} {\bibinfo {author} {\bibfnamefont {T.}~\bibnamefont
			{Geballe}}\ and\ \bibinfo {author} {\bibfnamefont {G.}~\bibnamefont {Hull}},\
	}\bibfield  {title} {\bibinfo {title} {Seebeck effect in silicon},\
	}\href@noop {} {\bibfield  {journal} {\bibinfo  {journal} {Physical Review}\
		}\textbf {\bibinfo {volume} {98}},\ \bibinfo {pages} {940} (\bibinfo {year}
		{1955})}\BibitemShut {NoStop}%
	\bibitem [{\citenamefont {Herring}(1954)}]{herring1954theory}%
	\BibitemOpen
	\bibfield  {author} {\bibinfo {author} {\bibfnamefont {C.}~\bibnamefont
			{Herring}},\ }\bibfield  {title} {\bibinfo {title} {Theory of the
			thermoelectric power of semiconductors},\ }\href@noop {} {\bibfield
		{journal} {\bibinfo  {journal} {Physical Review}\ }\textbf {\bibinfo {volume}
			{96}},\ \bibinfo {pages} {1163} (\bibinfo {year} {1954})}\BibitemShut
	{NoStop}%
	\bibitem [{\citenamefont {Cantrell}\ and\ \citenamefont
		{Butcher}(1987{\natexlab{a}})}]{cantrell1987calculation1}%
	\BibitemOpen
	\bibfield  {author} {\bibinfo {author} {\bibfnamefont {D.}~\bibnamefont
			{Cantrell}}\ and\ \bibinfo {author} {\bibfnamefont {P.}~\bibnamefont
			{Butcher}},\ }\bibfield  {title} {\bibinfo {title} {{A calculation of the
				phonon-drag contribution to the thermopower of quasi-2D electrons coupled to
				3D phonons. I. General theory}},\ }\href@noop {} {\bibfield  {journal}
		{\bibinfo  {journal} {Journal of Physics C: Solid State Physics}\ }\textbf
		{\bibinfo {volume} {20}},\ \bibinfo {pages} {1985} (\bibinfo {year}
		{1987}{\natexlab{a}})}\BibitemShut {NoStop}%
	\bibitem [{\citenamefont {Cantrell}\ and\ \citenamefont
		{Butcher}(1987{\natexlab{b}})}]{cantrell1987calculation2}%
	\BibitemOpen
	\bibfield  {author} {\bibinfo {author} {\bibfnamefont {D.}~\bibnamefont
			{Cantrell}}\ and\ \bibinfo {author} {\bibfnamefont {P.}~\bibnamefont
			{Butcher}},\ }\bibfield  {title} {\bibinfo {title} {{A calculation of the
				phonon-drag contribution to the thermopower of quasi-2D electrons coupled to
				3D phonons. II. Applications}},\ }\href@noop {} {\bibfield  {journal}
		{\bibinfo  {journal} {Journal of Physics C: Solid State Physics}\ }\textbf
		{\bibinfo {volume} {20}},\ \bibinfo {pages} {1993} (\bibinfo {year}
		{1987}{\natexlab{b}})}\BibitemShut {NoStop}%
	\bibitem [{\citenamefont {Tsaousidou}\ \emph {et~al.}(2001)\citenamefont
		{Tsaousidou}, \citenamefont {Butcher},\ and\ \citenamefont
		{Triberis}}]{tsaousidou2001fundamental}%
	\BibitemOpen
	\bibfield  {author} {\bibinfo {author} {\bibfnamefont {M.}~\bibnamefont
			{Tsaousidou}}, \bibinfo {author} {\bibfnamefont {P.}~\bibnamefont
			{Butcher}},\ and\ \bibinfo {author} {\bibfnamefont {G.}~\bibnamefont
			{Triberis}},\ }\bibfield  {title} {\bibinfo {title} {{Fundamental
				relationship between the Herring and Cantrell-Butcher formulas for the
				phonon-drag thermopower of two-dimensional electron and hole gases}},\
	}\href@noop {} {\bibfield  {journal} {\bibinfo  {journal} {Physical Review
				B}\ }\textbf {\bibinfo {volume} {64}},\ \bibinfo {pages} {165304} (\bibinfo
		{year} {2001})}\BibitemShut {NoStop}%
	\bibitem [{\citenamefont {Mahan}\ \emph {et~al.}(2014)\citenamefont {Mahan},
		\citenamefont {Lindsay},\ and\ \citenamefont {Broido}}]{mahan2014seebeck}%
	\BibitemOpen
	\bibfield  {author} {\bibinfo {author} {\bibfnamefont {G.}~\bibnamefont
			{Mahan}}, \bibinfo {author} {\bibfnamefont {L.}~\bibnamefont {Lindsay}},\
		and\ \bibinfo {author} {\bibfnamefont {D.}~\bibnamefont {Broido}},\
	}\bibfield  {title} {\bibinfo {title} {{The Seebeck coefficient and phonon
				drag in silicon}},\ }\href@noop {} {\bibfield  {journal} {\bibinfo  {journal}
			{Journal of Applied Physics}\ }\textbf {\bibinfo {volume} {116}},\ \bibinfo
		{pages} {245102} (\bibinfo {year} {2014})}\BibitemShut {NoStop}%
	\bibitem [{\citenamefont {Rode}(1970)}]{rode1970electron}%
	\BibitemOpen
	\bibfield  {author} {\bibinfo {author} {\bibfnamefont {D.}~\bibnamefont
			{Rode}},\ }\bibfield  {title} {\bibinfo {title} {Electron mobility in
			direct-gap polar semiconductors},\ }\href@noop {} {\bibfield  {journal}
		{\bibinfo  {journal} {Physical Review B}\ }\textbf {\bibinfo {volume} {2}},\
		\bibinfo {pages} {1012} (\bibinfo {year} {1970})}\BibitemShut {NoStop}%
	\bibitem [{\citenamefont {Zhou}\ \emph {et~al.}(2015)\citenamefont {Zhou},
		\citenamefont {Liao}, \citenamefont {Qiu}, \citenamefont {Huberman},
		\citenamefont {Esfarjani}, \citenamefont {Dresselhaus},\ and\ \citenamefont
		{Chen}}]{zhou2015ab}%
	\BibitemOpen
	\bibfield  {author} {\bibinfo {author} {\bibfnamefont {J.}~\bibnamefont
			{Zhou}}, \bibinfo {author} {\bibfnamefont {B.}~\bibnamefont {Liao}}, \bibinfo
		{author} {\bibfnamefont {B.}~\bibnamefont {Qiu}}, \bibinfo {author}
		{\bibfnamefont {S.}~\bibnamefont {Huberman}}, \bibinfo {author}
		{\bibfnamefont {K.}~\bibnamefont {Esfarjani}}, \bibinfo {author}
		{\bibfnamefont {M.~S.}\ \bibnamefont {Dresselhaus}},\ and\ \bibinfo {author}
		{\bibfnamefont {G.}~\bibnamefont {Chen}},\ }\bibfield  {title} {\bibinfo
		{title} {Ab initio optimization of phonon drag effect for lower-temperature
			thermoelectric energy conversion},\ }\href@noop {} {\bibfield  {journal}
		{\bibinfo  {journal} {Proceedings of the National Academy of Sciences}\
		}\textbf {\bibinfo {volume} {112}},\ \bibinfo {pages} {14777} (\bibinfo
		{year} {2015})}\BibitemShut {NoStop}%
	\bibitem [{\citenamefont {Fiorentini}\ and\ \citenamefont
		{Bonini}(2016)}]{fiorentini2016thermoelectric}%
	\BibitemOpen
	\bibfield  {author} {\bibinfo {author} {\bibfnamefont {M.}~\bibnamefont
			{Fiorentini}}\ and\ \bibinfo {author} {\bibfnamefont {N.}~\bibnamefont
			{Bonini}},\ }\bibfield  {title} {\bibinfo {title} {{Thermoelectric
				coefficients of n-doped silicon from first principles via the solution of the
				Boltzmann transport equation}},\ }\href@noop {} {\bibfield  {journal}
		{\bibinfo  {journal} {Physical Review B}\ }\textbf {\bibinfo {volume} {94}},\
		\bibinfo {pages} {085204} (\bibinfo {year} {2016})}\BibitemShut {NoStop}%
	\bibitem [{\citenamefont {Macheda}\ and\ \citenamefont
		{Bonini}(2018)}]{macheda2018magnetotransport}%
	\BibitemOpen
	\bibfield  {author} {\bibinfo {author} {\bibfnamefont {F.}~\bibnamefont
			{Macheda}}\ and\ \bibinfo {author} {\bibfnamefont {N.}~\bibnamefont
			{Bonini}},\ }\bibfield  {title} {\bibinfo {title} {Magnetotransport phenomena
			in p-doped diamond from first principles},\ }\href@noop {} {\bibfield
		{journal} {\bibinfo  {journal} {Physical Review B}\ }\textbf {\bibinfo
			{volume} {98}},\ \bibinfo {pages} {201201} (\bibinfo {year}
		{2018})}\BibitemShut {NoStop}%
	\bibitem [{\citenamefont {Protik}\ and\ \citenamefont
		{Broido}(2020)}]{protik2020coupled}%
	\BibitemOpen
	\bibfield  {author} {\bibinfo {author} {\bibfnamefont {N.~H.}\ \bibnamefont
			{Protik}}\ and\ \bibinfo {author} {\bibfnamefont {D.~A.}\ \bibnamefont
			{Broido}},\ }\bibfield  {title} {\bibinfo {title} {{Coupled transport of
				phonons and carriers in semiconductors: A case study of \textit{n}-doped
				GaAs}},\ }\href@noop {} {\bibfield  {journal} {\bibinfo  {journal} {Physical
				Review B}\ }\textbf {\bibinfo {volume} {101}},\ \bibinfo {pages} {075202}
		(\bibinfo {year} {2020})}\BibitemShut {NoStop}%
	\bibitem [{\citenamefont {Ponc{\'e}}\ \emph {et~al.}(2016)\citenamefont
		{Ponc{\'e}}, \citenamefont {Margine}, \citenamefont {Verdi},\ and\
		\citenamefont {Giustino}}]{ponce2016epw}%
	\BibitemOpen
	\bibfield  {author} {\bibinfo {author} {\bibfnamefont {S.}~\bibnamefont
			{Ponc{\'e}}}, \bibinfo {author} {\bibfnamefont {E.~R.}\ \bibnamefont
			{Margine}}, \bibinfo {author} {\bibfnamefont {C.}~\bibnamefont {Verdi}},\
		and\ \bibinfo {author} {\bibfnamefont {F.}~\bibnamefont {Giustino}},\
	}\bibfield  {title} {\bibinfo {title} {{EPW: Electron--phonon coupling,
				transport and superconducting properties using maximally localized Wannier
				functions}},\ }\href@noop {} {\bibfield  {journal} {\bibinfo  {journal}
			{Computer Physics Communications}\ }\textbf {\bibinfo {volume} {209}},\
		\bibinfo {pages} {116} (\bibinfo {year} {2016})}\BibitemShut {NoStop}%
	\bibitem [{\citenamefont {Li}\ \emph {et~al.}(2014)\citenamefont {Li},
		\citenamefont {Carrete}, \citenamefont {Katcho},\ and\ \citenamefont
		{Mingo}}]{li2014shengbte}%
	\BibitemOpen
	\bibfield  {author} {\bibinfo {author} {\bibfnamefont {W.}~\bibnamefont
			{Li}}, \bibinfo {author} {\bibfnamefont {J.}~\bibnamefont {Carrete}},
		\bibinfo {author} {\bibfnamefont {N.~A.}\ \bibnamefont {Katcho}},\ and\
		\bibinfo {author} {\bibfnamefont {N.}~\bibnamefont {Mingo}},\ }\bibfield
	{title} {\bibinfo {title} {{ShengBTE: A solver of the Boltzmann transport
				equation for phonons}},\ }\href@noop {} {\bibfield  {journal} {\bibinfo
			{journal} {Computer Physics Communications}\ }\textbf {\bibinfo {volume}
			{185}},\ \bibinfo {pages} {1747} (\bibinfo {year} {2014})}\BibitemShut
	{NoStop}%
	\bibitem [{\citenamefont {Giannozzi}\ \emph {et~al.}(2017)\citenamefont
		{Giannozzi}, \citenamefont {Andreussi}, \citenamefont {Brumme}, \citenamefont
		{Bunau}, \citenamefont {Nardelli}, \citenamefont {Calandra}, \citenamefont
		{Car}, \citenamefont {Cavazzoni}, \citenamefont {Ceresoli}, \citenamefont
		{Cococcioni} \emph {et~al.}}]{giannozzi2017advanced}%
	\BibitemOpen
	\bibfield  {author} {\bibinfo {author} {\bibfnamefont {P.}~\bibnamefont
			{Giannozzi}}, \bibinfo {author} {\bibfnamefont {O.}~\bibnamefont
			{Andreussi}}, \bibinfo {author} {\bibfnamefont {T.}~\bibnamefont {Brumme}},
		\bibinfo {author} {\bibfnamefont {O.}~\bibnamefont {Bunau}}, \bibinfo
		{author} {\bibfnamefont {M.~B.}\ \bibnamefont {Nardelli}}, \bibinfo {author}
		{\bibfnamefont {M.}~\bibnamefont {Calandra}}, \bibinfo {author}
		{\bibfnamefont {R.}~\bibnamefont {Car}}, \bibinfo {author} {\bibfnamefont
			{C.}~\bibnamefont {Cavazzoni}}, \bibinfo {author} {\bibfnamefont
			{D.}~\bibnamefont {Ceresoli}}, \bibinfo {author} {\bibfnamefont
			{M.}~\bibnamefont {Cococcioni}}, \emph {et~al.},\ }\bibfield  {title}
	{\bibinfo {title} {Advanced capabilities for materials modelling with quantum
			espresso},\ }\href@noop {} {\bibfield  {journal} {\bibinfo  {journal}
			{Journal of Physics: Condensed Matter}\ }\textbf {\bibinfo {volume} {29}},\
		\bibinfo {pages} {465901} (\bibinfo {year} {2017})}\BibitemShut {NoStop}%
	\bibitem [{\citenamefont {Perdew}\ and\ \citenamefont
		{Zunger}(1981)}]{perdew1981self}%
	\BibitemOpen
	\bibfield  {author} {\bibinfo {author} {\bibfnamefont {J.~P.}\ \bibnamefont
			{Perdew}}\ and\ \bibinfo {author} {\bibfnamefont {A.}~\bibnamefont
			{Zunger}},\ }\bibfield  {title} {\bibinfo {title} {Self-interaction
			correction to density-functional approximations for many-electron systems},\
	}\href@noop {} {\bibfield  {journal} {\bibinfo  {journal} {Physical Review
				B}\ }\textbf {\bibinfo {volume} {23}},\ \bibinfo {pages} {5048} (\bibinfo
		{year} {1981})}\BibitemShut {NoStop}%
	\bibitem [{\citenamefont {Smiltens}(1960)}]{smiltens1960silicon}%
	\BibitemOpen
	\bibfield  {author} {\bibinfo {author} {\bibfnamefont {J.}~\bibnamefont
			{Smiltens}},\ }\href@noop {} {\emph {\bibinfo {title} {Silicon Carbide: A
				High Temperature Semiconductor: Proceedings of the Conference on Silicon
				Carbide, April, 2-3, 1959, Boston, Massachusetts}}}\ (\bibinfo  {publisher}
	{Pergamon Press},\ \bibinfo {year} {1960})\BibitemShut {NoStop}%
	\bibitem [{\citenamefont {Lindsay}\ \emph {et~al.}(2013)\citenamefont
		{Lindsay}, \citenamefont {Broido},\ and\ \citenamefont
		{Reinecke}}]{lindsay2013ab}%
	\BibitemOpen
	\bibfield  {author} {\bibinfo {author} {\bibfnamefont {L.}~\bibnamefont
			{Lindsay}}, \bibinfo {author} {\bibfnamefont {D.}~\bibnamefont {Broido}},\
		and\ \bibinfo {author} {\bibfnamefont {T.}~\bibnamefont {Reinecke}},\
	}\bibfield  {title} {\bibinfo {title} {{Ab initio thermal transport in
				compound semiconductors}},\ }\href@noop {} {\bibfield  {journal} {\bibinfo
			{journal} {Physical Review B}\ }\textbf {\bibinfo {volume} {87}},\ \bibinfo
		{pages} {165201} (\bibinfo {year} {2013})}\BibitemShut {NoStop}%
	\bibitem [{\citenamefont {Katre}\ \emph {et~al.}(2017)\citenamefont {Katre},
		\citenamefont {Carrete}, \citenamefont {Dongre}, \citenamefont {Madsen},\
		and\ \citenamefont {Mingo}}]{katre2017exceptionally}%
	\BibitemOpen
	\bibfield  {author} {\bibinfo {author} {\bibfnamefont {A.}~\bibnamefont
			{Katre}}, \bibinfo {author} {\bibfnamefont {J.}~\bibnamefont {Carrete}},
		\bibinfo {author} {\bibfnamefont {B.}~\bibnamefont {Dongre}}, \bibinfo
		{author} {\bibfnamefont {G.~K.}\ \bibnamefont {Madsen}},\ and\ \bibinfo
		{author} {\bibfnamefont {N.}~\bibnamefont {Mingo}},\ }\bibfield  {title}
	{\bibinfo {title} {{Exceptionally strong phonon scattering by B substitution
				in cubic SiC}},\ }\href@noop {} {\bibfield  {journal} {\bibinfo  {journal}
			{Physical Review Letters}\ }\textbf {\bibinfo {volume} {119}},\ \bibinfo
		{pages} {075902} (\bibinfo {year} {2017})}\BibitemShut {NoStop}%
	\bibitem [{\citenamefont {Wang}\ \emph {et~al.}(2017)\citenamefont {Wang},
		\citenamefont {Gui}, \citenamefont {Janotti}, \citenamefont {Ni},\ and\
		\citenamefont {Karandikar}}]{wang2017strong}%
	\BibitemOpen
	\bibfield  {author} {\bibinfo {author} {\bibfnamefont {T.}~\bibnamefont
			{Wang}}, \bibinfo {author} {\bibfnamefont {Z.}~\bibnamefont {Gui}}, \bibinfo
		{author} {\bibfnamefont {A.}~\bibnamefont {Janotti}}, \bibinfo {author}
		{\bibfnamefont {C.}~\bibnamefont {Ni}},\ and\ \bibinfo {author}
		{\bibfnamefont {P.}~\bibnamefont {Karandikar}},\ }\bibfield  {title}
	{\bibinfo {title} {{Strong effect of electron-phonon interaction on the
				lattice thermal conductivity in 3C-SiC}},\ }\href@noop {} {\bibfield
		{journal} {\bibinfo  {journal} {Physical Review Materials}\ }\textbf
		{\bibinfo {volume} {1}},\ \bibinfo {pages} {034601} (\bibinfo {year}
		{2017})}\BibitemShut {NoStop}%
	\bibitem [{\citenamefont {Giustino}\ \emph {et~al.}(2007)\citenamefont
		{Giustino}, \citenamefont {Cohen},\ and\ \citenamefont
		{Louie}}]{giustino2007electron}%
	\BibitemOpen
	\bibfield  {author} {\bibinfo {author} {\bibfnamefont {F.}~\bibnamefont
			{Giustino}}, \bibinfo {author} {\bibfnamefont {M.~L.}\ \bibnamefont
			{Cohen}},\ and\ \bibinfo {author} {\bibfnamefont {S.~G.}\ \bibnamefont
			{Louie}},\ }\bibfield  {title} {\bibinfo {title} {{Electron-phonon
				interaction using Wannier functions}},\ }\href@noop {} {\bibfield  {journal}
		{\bibinfo  {journal} {Physical Review B}\ }\textbf {\bibinfo {volume} {76}},\
		\bibinfo {pages} {165108} (\bibinfo {year} {2007})}\BibitemShut {NoStop}%
	\bibitem [{\citenamefont {Verdi}\ and\ \citenamefont
		{Giustino}(2015)}]{verdi2015frohlich}%
	\BibitemOpen
	\bibfield  {author} {\bibinfo {author} {\bibfnamefont {C.}~\bibnamefont
			{Verdi}}\ and\ \bibinfo {author} {\bibfnamefont {F.}~\bibnamefont
			{Giustino}},\ }\bibfield  {title} {\bibinfo {title} {Fr{\"o}hlich
			electron-phonon vertex from first principles},\ }\href@noop {} {\bibfield
		{journal} {\bibinfo  {journal} {Physical Review Letters}\ }\textbf {\bibinfo
			{volume} {115}},\ \bibinfo {pages} {176401} (\bibinfo {year}
		{2015})}\BibitemShut {NoStop}%
	\bibitem [{\citenamefont {Lambin}\ and\ \citenamefont
		{Vigneron}(1984)}]{lambin1984computation}%
	\BibitemOpen
	\bibfield  {author} {\bibinfo {author} {\bibfnamefont {P.}~\bibnamefont
			{Lambin}}\ and\ \bibinfo {author} {\bibfnamefont {J.-P.}\ \bibnamefont
			{Vigneron}},\ }\bibfield  {title} {\bibinfo {title} {Computation of crystal
			green's functions in the complex-energy plane with the use of the analytical
			tetrahedron method},\ }\href@noop {} {\bibfield  {journal} {\bibinfo
			{journal} {Physical Review B}\ }\textbf {\bibinfo {volume} {29}},\ \bibinfo
		{pages} {3430} (\bibinfo {year} {1984})}\BibitemShut {NoStop}%
	\bibitem [{\citenamefont {Chattopadhyay}\ and\ \citenamefont
		{Queisser}(1981)}]{chattopadhyay1981electron}%
	\BibitemOpen
	\bibfield  {author} {\bibinfo {author} {\bibfnamefont {D.}~\bibnamefont
			{Chattopadhyay}}\ and\ \bibinfo {author} {\bibfnamefont {H.}~\bibnamefont
			{Queisser}},\ }\bibfield  {title} {\bibinfo {title} {Electron scattering by
			ionized impurities in semiconductors},\ }\href@noop {} {\bibfield  {journal}
		{\bibinfo  {journal} {Reviews of Modern Physics}\ }\textbf {\bibinfo {volume}
			{53}},\ \bibinfo {pages} {745} (\bibinfo {year} {1981})}\BibitemShut
	{NoStop}%
	\bibitem [{\citenamefont {Li}\ \emph {et~al.}(2012)\citenamefont {Li},
		\citenamefont {Lindsay}, \citenamefont {Broido}, \citenamefont {Stewart},\
		and\ \citenamefont {Mingo}}]{li2012thermal}%
	\BibitemOpen
	\bibfield  {author} {\bibinfo {author} {\bibfnamefont {W.}~\bibnamefont
			{Li}}, \bibinfo {author} {\bibfnamefont {L.}~\bibnamefont {Lindsay}},
		\bibinfo {author} {\bibfnamefont {D.}~\bibnamefont {Broido}}, \bibinfo
		{author} {\bibfnamefont {D.~A.}\ \bibnamefont {Stewart}},\ and\ \bibinfo
		{author} {\bibfnamefont {N.}~\bibnamefont {Mingo}},\ }\bibfield  {title}
	{\bibinfo {title} {{Thermal conductivity of bulk and nanowire
				Mg$_{2}$Si$_{x}$Sn$_{1- x}$ alloys from first principles}},\ }\href@noop {}
	{\bibfield  {journal} {\bibinfo  {journal} {Physical Review B}\ }\textbf
		{\bibinfo {volume} {86}},\ \bibinfo {pages} {174307} (\bibinfo {year}
		{2012})}\BibitemShut {NoStop}%
	\bibitem [{\citenamefont {Tamura}(1983)}]{tamura1983isotope}%
	\BibitemOpen
	\bibfield  {author} {\bibinfo {author} {\bibfnamefont {S.-i.}\ \bibnamefont
			{Tamura}},\ }\bibfield  {title} {\bibinfo {title} {Isotope scattering of
			dispersive phonons in ge},\ }\href@noop {} {\bibfield  {journal} {\bibinfo
			{journal} {Physical Review B}\ }\textbf {\bibinfo {volume} {27}},\ \bibinfo
		{pages} {858} (\bibinfo {year} {1983})}\BibitemShut {NoStop}%
	\bibitem [{\citenamefont {Patrick}(1966)}]{patrick1966high}%
	\BibitemOpen
	\bibfield  {author} {\bibinfo {author} {\bibfnamefont {L.}~\bibnamefont
			{Patrick}},\ }\bibfield  {title} {\bibinfo {title} {{High Electron Mobility
				of Cubic SiC}},\ }\href@noop {} {\bibfield  {journal} {\bibinfo  {journal}
			{Journal of Applied Physics}\ }\textbf {\bibinfo {volume} {37}},\ \bibinfo
		{pages} {4911} (\bibinfo {year} {1966})}\BibitemShut {NoStop}%
	\bibitem [{\citenamefont {Meng}\ \emph {et~al.}(2019)\citenamefont {Meng},
		\citenamefont {Ma}, \citenamefont {He},\ and\ \citenamefont
		{Li}}]{meng2019phonon}%
	\BibitemOpen
	\bibfield  {author} {\bibinfo {author} {\bibfnamefont {F.}~\bibnamefont
			{Meng}}, \bibinfo {author} {\bibfnamefont {J.}~\bibnamefont {Ma}}, \bibinfo
		{author} {\bibfnamefont {J.}~\bibnamefont {He}},\ and\ \bibinfo {author}
		{\bibfnamefont {W.}~\bibnamefont {Li}},\ }\bibfield  {title} {\bibinfo
		{title} {{Phonon-limited carrier mobility and temperature-dependent
				scattering mechanism of 3C-SiC from first principles}},\ }\href@noop {}
	{\bibfield  {journal} {\bibinfo  {journal} {Physical Review B}\ }\textbf
		{\bibinfo {volume} {99}},\ \bibinfo {pages} {045201} (\bibinfo {year}
		{2019})}\BibitemShut {NoStop}%
	\bibitem [{\citenamefont {Nelson}\ \emph {et~al.}(1966)\citenamefont {Nelson},
		\citenamefont {Halden},\ and\ \citenamefont {Rosengreen}}]{nelson1966growth}%
	\BibitemOpen
	\bibfield  {author} {\bibinfo {author} {\bibfnamefont {W.}~\bibnamefont
			{Nelson}}, \bibinfo {author} {\bibfnamefont {F.}~\bibnamefont {Halden}},\
		and\ \bibinfo {author} {\bibfnamefont {A.}~\bibnamefont {Rosengreen}},\
	}\bibfield  {title} {\bibinfo {title} {{Growth and Properties of $\beta$-SiC
				Single Crystals}},\ }\href@noop {} {\bibfield  {journal} {\bibinfo  {journal}
			{Journal of Applied Physics}\ }\textbf {\bibinfo {volume} {37}},\ \bibinfo
		{pages} {333} (\bibinfo {year} {1966})}\BibitemShut {NoStop}%
	\bibitem [{\citenamefont {Thesberg}\ \emph {et~al.}(2017)\citenamefont
		{Thesberg}, \citenamefont {Kosina},\ and\ \citenamefont
		{Neophytou}}]{thesberg2017lorenz}%
	\BibitemOpen
	\bibfield  {author} {\bibinfo {author} {\bibfnamefont {M.}~\bibnamefont
			{Thesberg}}, \bibinfo {author} {\bibfnamefont {H.}~\bibnamefont {Kosina}},\
		and\ \bibinfo {author} {\bibfnamefont {N.}~\bibnamefont {Neophytou}},\
	}\bibfield  {title} {\bibinfo {title} {{On the Lorenz number of multiband
				materials}},\ }\href@noop {} {\bibfield  {journal} {\bibinfo  {journal}
			{Physical Review B}\ }\textbf {\bibinfo {volume} {95}},\ \bibinfo {pages}
		{125206} (\bibinfo {year} {2017})}\BibitemShut {NoStop}%
	\bibitem [{\citenamefont {Putatunda}\ and\ \citenamefont
		{Singh}(2019)}]{putatunda2019lorenz}%
	\BibitemOpen
	\bibfield  {author} {\bibinfo {author} {\bibfnamefont {A.}~\bibnamefont
			{Putatunda}}\ and\ \bibinfo {author} {\bibfnamefont {D.}~\bibnamefont
			{Singh}},\ }\bibfield  {title} {\bibinfo {title} {{Lorenz number in relation
				to estimates based on the Seebeck coefficient}},\ }\href@noop {} {\bibfield
		{journal} {\bibinfo  {journal} {Materials Today Physics}\ }\textbf {\bibinfo
			{volume} {8}},\ \bibinfo {pages} {49} (\bibinfo {year} {2019})}\BibitemShut
	{NoStop}%
	\bibitem [{\citenamefont {Sondheimer}(1956)}]{sondheimer1956kelvin}%
	\BibitemOpen
	\bibfield  {author} {\bibinfo {author} {\bibfnamefont {E.}~\bibnamefont
			{Sondheimer}},\ }\bibfield  {title} {\bibinfo {title} {{The Kelvin relations
				in thermo-electricity}},\ }\href@noop {} {\bibfield  {journal} {\bibinfo
			{journal} {Proceedings of the Royal Society of London. Series A. Mathematical
				and Physical Sciences}\ }\textbf {\bibinfo {volume} {234}},\ \bibinfo {pages}
		{391} (\bibinfo {year} {1956})}\BibitemShut {NoStop}%
	\bibitem [{\citenamefont {Nag}(2012)}]{nag2012electron}%
	\BibitemOpen
	\bibfield  {author} {\bibinfo {author} {\bibfnamefont {B.~R.}\ \bibnamefont
			{Nag}},\ }\href@noop {} {\emph {\bibinfo {title} {Electron transport in
				compound semiconductors}}},\ Vol.~\bibinfo {volume} {11}\ (\bibinfo
	{publisher} {Springer Science \& Business Media},\ \bibinfo {year}
	{2012})\BibitemShut {NoStop}%
	\bibitem [{\citenamefont {Sanborn}\ \emph {et~al.}(1992)\citenamefont
		{Sanborn}, \citenamefont {Allen},\ and\ \citenamefont
		{Mahan}}]{sanborn1992theory}%
	\BibitemOpen
	\bibfield  {author} {\bibinfo {author} {\bibfnamefont {B.}~\bibnamefont
			{Sanborn}}, \bibinfo {author} {\bibfnamefont {P.}~\bibnamefont {Allen}},\
		and\ \bibinfo {author} {\bibfnamefont {G.}~\bibnamefont {Mahan}},\ }\bibfield
	{title} {\bibinfo {title} {Theory of screening and electron mobility:
			Application to n-type silicon},\ }\href@noop {} {\bibfield  {journal}
		{\bibinfo  {journal} {Physical Review B}\ }\textbf {\bibinfo {volume} {46}},\
		\bibinfo {pages} {15123} (\bibinfo {year} {1992})}\BibitemShut {NoStop}%
	\bibitem [{\citenamefont {Kaasbjerg}\ \emph {et~al.}(2012)\citenamefont
		{Kaasbjerg}, \citenamefont {Thygesen},\ and\ \citenamefont
		{Jacobsen}}]{kaasbjerg2012phonon}%
	\BibitemOpen
	\bibfield  {author} {\bibinfo {author} {\bibfnamefont {K.}~\bibnamefont
			{Kaasbjerg}}, \bibinfo {author} {\bibfnamefont {K.~S.}\ \bibnamefont
			{Thygesen}},\ and\ \bibinfo {author} {\bibfnamefont {K.~W.}\ \bibnamefont
			{Jacobsen}},\ }\bibfield  {title} {\bibinfo {title} {{Phonon-limited mobility
				in n-type single-layer MoS$_{2}$ from first principles}},\ }\href@noop {}
	{\bibfield  {journal} {\bibinfo  {journal} {Physical Review B}\ }\textbf
		{\bibinfo {volume} {85}},\ \bibinfo {pages} {115317} (\bibinfo {year}
		{2012})}\BibitemShut {NoStop}%
	\bibitem [{\citenamefont {{Ioffe database}}()}]{sicioffe}%
	\BibitemOpen
	\bibfield  {author} {\bibinfo {author} {\bibnamefont {{Ioffe database}}},\
	}\href@noop {} {\bibinfo {title} {{SiC materials parameters}}},\ \bibinfo
	{note} {\url{http://www.ioffe.ru/SVA/NSM/Semicond/SiC/}}\BibitemShut
	{NoStop}%
	\bibitem [{\citenamefont {{Materials Project database}}()}]{sicmatproj}%
	\BibitemOpen
	\bibfield  {author} {\bibinfo {author} {\bibnamefont {{Materials Project
					database}}},\ }\href@noop {} {\bibinfo {title} {{SiC materials
				paramaters}}},\ \bibinfo {note}
	{\url{https://materialsproject.org/materials/mp-8062/}}\BibitemShut {NoStop}%
	\bibitem [{\citenamefont {Liu}\ \emph {et~al.}(2017)\citenamefont {Liu},
		\citenamefont {Zhou}, \citenamefont {Liao}, \citenamefont {Singh},\ and\
		\citenamefont {Chen}}]{liu2017first}%
	\BibitemOpen
	\bibfield  {author} {\bibinfo {author} {\bibfnamefont {T.-H.}\ \bibnamefont
			{Liu}}, \bibinfo {author} {\bibfnamefont {J.}~\bibnamefont {Zhou}}, \bibinfo
		{author} {\bibfnamefont {B.}~\bibnamefont {Liao}}, \bibinfo {author}
		{\bibfnamefont {D.~J.}\ \bibnamefont {Singh}},\ and\ \bibinfo {author}
		{\bibfnamefont {G.}~\bibnamefont {Chen}},\ }\bibfield  {title} {\bibinfo
		{title} {{First-principles mode-by-mode analysis for electron-phonon
				scattering channels and mean free path spectra in GaAs}},\ }\href@noop {}
	{\bibfield  {journal} {\bibinfo  {journal} {Physical Review B}\ }\textbf
		{\bibinfo {volume} {95}},\ \bibinfo {pages} {075206} (\bibinfo {year}
		{2017})}\BibitemShut {NoStop}%
	\bibitem [{\citenamefont {Jacoboni}\ \emph {et~al.}(1977)\citenamefont
		{Jacoboni}, \citenamefont {Canali}, \citenamefont {Ottaviani},\ and\
		\citenamefont {Quaranta}}]{jacoboni1977review}%
	\BibitemOpen
	\bibfield  {author} {\bibinfo {author} {\bibfnamefont {C.}~\bibnamefont
			{Jacoboni}}, \bibinfo {author} {\bibfnamefont {C.}~\bibnamefont {Canali}},
		\bibinfo {author} {\bibfnamefont {G.}~\bibnamefont {Ottaviani}},\ and\
		\bibinfo {author} {\bibfnamefont {A.~A.}\ \bibnamefont {Quaranta}},\
	}\bibfield  {title} {\bibinfo {title} {A review of some charge transport
			properties of silicon},\ }\href@noop {} {\bibfield  {journal} {\bibinfo
			{journal} {Solid-State Electronics}\ }\textbf {\bibinfo {volume} {20}},\
		\bibinfo {pages} {77} (\bibinfo {year} {1977})}\BibitemShut {NoStop}%
\end{thebibliography}
%apsrev4-2.bst 2019-01-14 (MD) hand-edited version of apsrev4-1.bst
%Control: key (0)
%Control: author (8) initials jnrlst
%Control: editor formatted (1) identically to author
%Control: production of article title (0) allowed
%Control: page (0) single
%Control: year (1) truncated
%Control: production of eprint (0) enabled
%

\end{document}